\documentstyle[12pt,epsf]{article}
\begin{document}
\baselineskip=22.pt plus 0.2pt minus 0.2pt
\lineskip=22.pt plus 0.2pt minus 0.2pt

\setcounter{equation}{0}

\centerline{\Large\bf Model Independent Features of}
\centerline{\Large\bf the Two-Particle Correlation Function}

\vskip 0.8cm

\centerline{\large Scott Chapman, Pierre Scotto and Ulrich Heinz}

\centerline{\em Institut f\"ur Theoretische Physik, Universit\"at
Regensburg,}
\centerline{\em D-93040 Regensburg, Germany}

\vskip 0.8cm
\noindent\underbar{\bf Abstract:}
The Hanbury-Brown Twiss correlation function for two identical
particles is studied for systems with cylindrical symmetry. Its shape
for small values of the relative momentum is derived in a model
independent way. In addition to the usual quadratic ``side'', ``out''
and ``longitudinal'' terms in the exponent of the correlator, a
previously neglected ``out-longitudinal'' cross term is found and
discussed.  The model-independent expressions for the size parameters
of the HBT correlation function are interpreted as lengths of
homogeneity of the source, in distinction to its purely geometrical
size.  They are evaluated analytically and numerically for two
specific thermal models featuring collective transverse and
longitudinal flow.  The analytic expressions derived allow one to
establish qualitatively important connections between the space-time
features of the source and the shape of the correlation function.  New
ways of parametrizing the correlation function and a new approach to
the measurement of the duration of the emission process are suggested.

\newpage

\section{Introduction}

It is widely accepted that if the nuclear matter created in
ultra-relativistic heavy-ion collisions attains a high enough energy
density, it will undergo a phase transition into a quark-gluon plasma.
For this reason, it is of great interest to determine the energy
densities actually attained in these collisions. The total interaction
energy of a given reaction can be directly measured by particle
calorimeters and spectrometers. Although there is no analogous direct
measurement for the size of the reaction region, Hanbury-Brown Twiss
(HBT) interferometry \cite{hbt} provides an indirect measurement for both
the spatial and temporal extent of the reaction region in terms of the
correlations between produced particles.

Consequently, the greatest challenge for theorists studying HBT
interferometry today is to determine exactly what information the
reported experimental correlation radii are telling us about the
source.  Obviously, the most powerful statements to this effect are
those which can be made in a model-independent fashion.  Although the
individual reactions measured experimentally may not be completely
cylindrically symmetric, it is safe to assume that a large ensemble of
similar reactions will produce cylindrically symmetric data.  For this
reason, we have generalized the work of \cite{bertsch} by using the
covariant Wigner function formulation [2-6] of HBT interferometry to
derive cylindrically symmetric, but otherwise model independent
expressions for the correlation radii, both using standard cartesian
momentum differences and boost-invariant rapidity differences.  Two
important model-independent statements can then be made.  First,
cylindrical symmetry in no way precludes the existence of an
``out-longitudinal'' cross term in the correlation function
\cite{chaplett}, and in fact in general such a term would be expected
to appear.  Second, the correlation radii do not necessarily measure
the geometrical size of the reaction region, but rather the lengths of
homogeneity of the source as seen by a particle emitted with the
average momentum of the studied pair \cite{sinyu3}.

To see how these effects manifest themselves in a concrete (though
still qualitative) way, we apply our model-independent formalism to
two specific thermal models, both of which feature a constant
freezeout temperature.  The first model is a generalization of
\cite{csorgo}, featuring nonrelativistic hydrodynamic flow which,
however, can be different in the longitudinal and the transversal
directions.  Since this model is completely gaussian, it is easy to
verify explicitly that the spatial lengths of homogeneity depend not
only on the geometrical size of the reaction region, but also on the
spatial gradients of the hydrodynamic flow.  Similarly, the cross term
just measures the temporal length of homogeneity, which in this
nonrelativistic case is simply the duration of particle emission.

The second model that we consider is a variation of \cite{csorgo2},
featuring a Bjorken scaling longitudinal flow and a nonrelativistic
transverse flow.  Although this model is not completely gaussian,
analytic results derived from a modified saddle point approximation
are able to reproduce numerically generated results to within 20-30\%
for pions and much better for kaons.  The analytic results provide
valuable qualitative insights into the generic influence of various
physically relevant parameters of the source distribution on the shape
of the correlation function.  We show that this model features a large
cross term whose effects can clearly be seen in a two-dimensional plot
of the ``out-longitudinal'' correlation function.  In addition, we
show that the theoretical interpretation of the correlation radii
simplifies immensely when rapidity differences rather than
longitudinal momentum differences are used to parametrize the
correlation functions.  In light of these results, we make explicit
suggestions of useful new ways in which experimentalists can organize
their measured correlation data.

\section{Model Independent Correlation Radii}

The HBT correlation function for two identical on-shell particles is
given by \cite{hbt,gyul}
\begin{equation}
C({\vec p}_1,{\vec p}_2) = \frac{\overline{N}^2}
{\overline{N^2}-\overline{N}}
\frac{P_2({\vec p}_1,{\vec p}_2)}{P_1({\vec p}_1)P_1({\vec p}_2)}\;,
\label{1}
\end{equation}
where $P_1({\vec p}\,)=E_p(dN/d^3p)$ is the invariant 1-particle
distribution for a particle with mass $m$ and 3-momentum ${\vec p}$,
$P_2$ is the corresponding invariant 2-particle distribution function,
and $\overline{N}$ ($\overline{N^2}$) is the average number of particles
(squared) produced in a reaction.  By quite general arguments it can
be shown that in the plane wave approximation for chaotic sources
[2-6]
\begin{equation}
   C({\vec p}_1,{\vec p}_2) = 1
   \pm \frac{\left|\int d^4x\, S[x\,,{\textstyle\frac{1}{2}}(p_1+p_2)]\,
                   e^{iq{\cdot}x}\right|^2}
   {P_1({\vec p}_1)P_1({\vec p}_2)}\;,
\label{2}
\end{equation}
where the $+$ ($-$) sign is for bosons (fermions), $q=p_1-p_2$ is the
4-momentum difference of the two particles, and $p_i^0=E_i$ are the
on-shell energies.  Furthermore, the emission function $S(x,p)$ is a
scalar function of the 4-vectors $x$ and $p$ which obeys
 \begin{equation}
   \int d^4x\, S(x,p_i) = P_1({\vec p_i})\;.
 \label{3}
\end{equation}

As an example, in the local hydrodynamic formulation involving a sharp
3-di\-men\-sio\-nal freeze-out hypersurface one has \cite{marb}
 \begin{equation}
   S(x,p) = \frac{1}{(2\pi)^3}
   \frac{p{\cdot}n(x)}{\exp[\beta(x)(p{\cdot}u(x)-\mu(x))]
   \mp1 } \;,
 \label{4}
\end{equation}
where $u_\mu(x)$, $\beta(x)$, $\mu(x)$ and
 \begin{equation}
   n_\mu(x)=
   \int_\Sigma d^3\sigma_\mu(x^\prime)\,\delta^{(4)}(x-x^\prime)
 \label{5}
 \end{equation}
denote the local hydrodynamic flow velocity, inverse temperature,
chemical potential, and normal-pointing freeze-out hypersurface
element, respectively.

\subsection{Cartesian Momentum Coordinates}
In order to simplify computation, the correlation function is
often approximated by using on-shell momenta in the emission function
\cite{bertsch,pratt,marb,sinyu}. For example, one can define \cite{pratt}
\begin{equation}
   C({\vec p}_1,{\vec p}_2) \simeq \tilde{C}
   ({\vec q},{\vec K}) =
   1 \pm \frac{|\int d^4x\, S(x,K)\,e^{iq{\cdot}x}|^2}
   {|\int d^4x\, S(x,K)|^2}
 \label{6}
\end{equation}
where ${\vec K} = {\textstyle\frac{1}{2}}({\vec p}_1 +{\vec p}_2)$ and
$K_0 = E_K =
\sqrt{m^2+|{\vec K}|^2}$.  Neither the present definition of $K$ nor
the different definition we will use in the next subsection should be
confused with the usual off-shell definition of
$K_0={\textstyle\frac{1}{2}}(E_1+E_2)$ which is suggested by
eq. (\ref{2}).

We begin by using the conventional HBT cartesian coordinate system
which is defined as follows: The ``longitudinal'' or $\hat{z}$
(subscript $L$) direction is defined to be parallel to the beam; the
``out'' or $\hat{x}$ (subscript $\perp$) direction is parallel to the
component of ${\vec K}$ which is perpendicular to the beam; and the
``side'' or $\hat{y}$ (subscript $s$) direction is the remaining
transverse direction.  For $|{\vec q}\,|/E_K \ll 1$, we then have
\begin{equation}
q \cdot x \simeq {\vec \beta}{\cdot}{\vec q}\,t
             - q_\perp\rho\cos\phi -q_s\rho\sin\phi -q_Lz
             \;,
\label{6a}
\end{equation}
where $\rho = \sqrt{x^2+y^2}$, $\phi=\tan^{-1}(y/x)$ and ${\vec
\beta}={\vec K}/E_K$ is the velocity of a particle with momentum
${\vec K}$.

To present their data, experimentalists use these coordinates in one
of two different reference frames, both of which can be obtained by a
longitudinal boost from the lab frame: The fixed observer frame is
usually taken as the rest frame of the participant center of mass and
is the same for all particle pairs \cite{na35,fixed,both}.  The ``LCMS''
(longitudinally co-moving system) frame, on the other hand, is defined
as the frame in which $K_L=0$ and thus varies for pairs with different
longitudinal momentum in the fixed observer frame \cite{nachtrag,both,lcms}.
Consequently, as pointed out in  \cite{nachtrag,csorgo,csorgo2}, a
$q_{L}$-correlation function should then only be measured at a given value of
$K_{L}$, and an averaging over $K_{L}$ should be avoided. However, since
different values of the longitudinal component of the mean momentum lead to
different reference frames, the interpretation of a possible $K_{L}$-dependence
of the correlation radii turns out to be conceptually nontrivial in the LCMS.
Later, however, we
will show that for the special case of a system which is undergoing
Bjorken longitudinal expansion, the LCMS radii are nothing more than
approximations of fixed frame radii which are evaluated in rapidity
coordinates (see next subsection).  To avoid the complication of shifting
reference frames, we perform all of our calculations in a fixed frame,
though we do point out how to find the LCMS results.

Due to the symmetry $C({\vec p}_1,{\vec p}_2)=C({\vec p}_2,{\vec
p}_1)$ and the fact that when $q\rightarrow 0$ the correlation
function $C({\vec p}_1,{\vec p}_2)\rightarrow 1\pm 1$ (as can be seen
from eq. (\ref{2})), it is reasonable to assume that for sufficiently
small momentum differences ${\vec q}$, $C$ takes the form
\begin{equation}
C({\vec p}_1,{\vec p}_2) = 1\pm \exp\left[
-\sum_i q_i^2R_i^2 - 2\sum_{i\ne j}q_iq_jR_{ij}^2\right]\;,
\label{6a1}
\end{equation}
where the coefficients $R_i^2$ and $R_{ij}^2$ depend on the average
pair momentum ${\vec K}={\textstyle\frac{1}{2}}({\vec p}_1+{\vec
p}_2)$.  Note that the $R_i^2$ are always positive, but the $R_{ij}^2$
can be either positive or negative; we simply use the $R_{ij}^2$
notation to denote the fact that they are coefficients of terms which
are quadratic in $q_i$.  Furthermore, in order for the peak of the
correlation function to be located at ${\vec q}=0$, it must be true
that for all $i$ and $j$
\begin{equation}
2|R_{ij}|^2 < R_i^2 + R_j^2\;.
\label{6a2}
\end{equation}
Below we obtain model independent expressions for the ``radii''
$R_i^2$ and $R_{ij}^2$ by effectively taking second derivatives of the
correlation function with respect to $q_i$ and $q_j$ around ${\vec
q}=0$.

Before proceeding, we would like to point out that one must take care
when comparing the above radii to experimentally measured correlation
radii since the former measure second derivatives of the correlation
function around ${\vec q}=0$, while the latter are parameters of a
gaussian fit to the whole correlation function
[14-17] and are essentially determined by its width.
Nevertheless, there are many interesting ``gaussian'' models for which
the two different ways of defining the radii give roughly the same
results.  To the extent that the part of the correlation function
measured by experimentalists is roughly gaussian, certain of these
``gaussian'' models should be able to provide good descriptions of the
data.  In this work we are therefore restricting the application of
our model independent results to ``gaussian'' models for which the
simple expressions that we generate below provide valuable insights as
to how various parameters of a given source distribution will
qualitatively affect measurable features of the correlation function.

Since $S(x,p)$ transforms as a scalar under Lorentz transformations,
it can be taken to have the following functional form
\begin{equation}
S(x,p) = \bar{S}(x,\,p{\cdot}p,\,p{\cdot}u(x),\,
p{\cdot}v(x),\,p{\cdot}w(x),...)\;,
\label{6b}
\end{equation}
where $u$, $v$, $w$, etc. are space-time dependent local 4-vectors.
Cylindrical symmetry can be enforced by demanding that $\bar{S}$ has
no explicit $\phi$ dependence and that all of the relevant local
4-vectors be cylindrically symmetric.  For example,
\begin{equation} u(x) = (u_0,u_\rho
\cos\phi, u_\rho \sin\phi, u_z)
\label{6c}
\end{equation}
where $u_0$, $u_\rho$ and $u_z$ are all
independent of $\phi$.  Given these definitions,
\begin{equation}
K{\cdot}u = E_Ku_0 - K_\perp u_\rho\cos\phi -K_L u_z\;,
\label{6d}
\end{equation}
so for cylindrically symmetric systems $S(x,K)$ is even in $\phi$.

Using (\ref{6a}), we can now expand the factor $\exp(iq{\cdot}x)$ in
eqn. (\ref{6}) for small ${\vec q}$, keeping only the terms even in
$\phi$, because the odd terms vanish upon $\phi$
integration. We find
\begin{eqnarray}
   \tilde{C}({\vec q},{\vec K}) = 1 \pm \Bigl\{1
   &-&
   q_s^2\langle y^2\rangle
   -\left\langle[q_\perp(x-\beta_\perp t) +
   q_L(z-\beta_L t)]^2\right\rangle
 \nonumber \\
   &+&
   \langle q_\perp(x-\beta_\perp t) +
   q_L(z-\beta_L t)\rangle^2
   + {\cal O}\left[\left\langle(q{\cdot}x)^4\right\rangle\right] \Bigr\} \, ,
\label{6e}
\end{eqnarray}
where $x=\rho\cos\phi$, $y=\rho\sin\phi$, and we have introduced the
notation
\begin{equation}
   \langle \xi\rangle =
   \frac{1}{P_1({\vec K})}\int d^4x\, \xi\, S(x,K)\;.
\label{6f}
\end{equation}
Eq. (\ref{6e}) generalizes similar results obtained in \cite{bertsch}
for a 1-dimensional situation.  Exponentiating (\ref{6e}), we can see
that for any cylindrically symmetric system the correlation function
for small momentum differences will take the form
\begin{equation}
   \tilde{C}({\vec q},{\vec K})
   \simeq 1 \pm \exp\left[ -q_s^2 R_s^2 - q_\perp^2 R_\perp^2
   - q_L^2 R_L^2 - 2q_\perp q_L R_{\perp L}^2\right]    \, .
\label{6g}
\end{equation}
The $R_i^2$ which correspond to the approximation (\ref{6}) can simply
be read off as the coefficients of the corresponding $q_iq_j$ terms in
eqn.~(\ref{6e}):
\begin{eqnarray}
R_s^2 & = & \left\langle y^2\right\rangle
\nonumber \\
R_\perp^2 & = &
\left\langle(x-\beta_\perp t)^2\right\rangle
  -\langle x-\beta_\perp t\rangle^2
\nonumber \\
R_L^2 & = &
\left\langle(z-\beta_L t)^2\right\rangle - \langle z-\beta_L t\rangle^2
\nonumber \\
R_{\perp L}^2 & = &
\left\langle(x-\beta_\perp t)(z-\beta_L t)\right\rangle
 - \langle x-\beta_\perp t\rangle \langle z-\beta_L t\rangle \;.
\label{6h}
\end{eqnarray}
They are functions of ${\vec K}$ due to the $K$-dependence of $S(x,K)$
in definition (\ref{6f}) of the expectation value $\langle ...
\rangle$.  One of the most
interesting features of (\ref{6g}) is, as pointed out in
\cite{chaplett}, the occurrence of a $q_\perp q_L$ cross term which
has never before been discussed in the literature.

Before exploring the implications of this term, we would like to
give an intuitive interpretation of the model-independent expressions
(\ref{6h}).  To this end we follow the work of
\cite{sinyu3} and introduce the concept of a length of homogeneity.
We begin by defining the spacetime saddle point $\bar{x}$ of the
emission function $S(x,K)$ through the four equations
\begin{equation}
\frac{d}{dx_\mu}\ln S(x,K) \Bigr|_{\bar{x}} = 0
\label{7a}
\end{equation}
where $\mu = \{0,1,2,3\}$.  Essentially the saddle point is that point
in space-time which has the maximum probability of emitting a particle
with momentum ${\vec K}$.  A saddle point approximation for $S(x,K)$
can then be made in the following way
\begin{equation}
S(x,K) \simeq S(\bar{x},K)\exp\left[-\sum_\mu
\frac{(x_\mu-\bar{x}_\mu)^2}{2\lambda_\mu^2}
-\sum_{\mu >\nu}
B_{\mu\nu}(x_\mu-\bar{x}_\mu)(x_\nu-\bar{x}_\nu)\right]\;,
\label{7b}
\end{equation}
where we define the length of homogeneity of the source in the $\mu$th
direction by
\begin{equation}
\lambda_\mu({\vec K}) =
\left[-\frac{d^2}{dx_\mu^2}\ln S(x,K)\Bigr|_{\bar{x}}
\right]^{-1/2}
\label{7c}
\end{equation}
and
\begin{equation}
B_{\mu\nu}({\vec K}) = -\frac{d}{dx_\mu}\frac{d}{dx_\nu}
\ln S(x,K)\Bigr|_{\bar{x}}\;.
\label{7d}
\end{equation}
{}From (\ref{7c}), it can be seen that the length of homogeneity
provides a measure of the region over which the source is relatively
constant as seen by a particle with momentum ${\vec K}$.  Obviously,
if a source has large temperature or flow gradients, the length of
homogeneity may be determined by these more than by geometrical
density gradients.

Notice that if $B_{\mu\nu}\ll 1/\lambda_\mu^2$ (which can always be
arranged by making the right choice of variables), then
\begin{equation}
\langle x_\mu^2\rangle - \langle x_\mu \rangle^2 \simeq \lambda_\mu^2
\label{7e}
\end{equation}
where we do not use the summation convention.  Since all of the radii
of eq.  (\ref{6h}) contain terms of the above form, {\em these radii
are evidently measuring lengths of homogeneity rather than strictly
geometrical sizes}.  For example, the ``side'' radius measures
$\lambda_2$ (in the $\hat{y}$ direction), which by cylindrical
symmetry must be equal to the length of homogeneity in the transverse
or radial direction.  Later, we will see how these lengths manifest
themselves in definite models.

Corrections to the radii of (\ref{6h}) can be calculated by
considering the exact correlation function (\ref{2}) rather than the
approximation (\ref{6}). (Within their one-dimensional model these
corrections were also found in \cite{bertsch}.)  The corrections to
the denominator can be found by noticing that due to cylindrical
symmetry, $P_1({\vec p}\,)$ is really only a function of the
longitudinal and radial components of ${\vec p}$. Hence,
\begin{eqnarray}
   P_1({\vec p}_1)  &=&
   \bar{P}_1\! \left( p_{1L},\, \sqrt{{\vec p}_1{\cdot}{\vec p}_1
                    - p_{1L}^2} \right)
\nonumber \\
   &=& \bar{P}_1\!  \left( K_L + {\textstyle\frac{1}{2}} q_L,\,
   K_\perp + {\textstyle\frac{1}{2}} q_\perp +\frac{q_s^2}{8K_\perp}
    +{\cal O}(|{\vec q}\,|^3/E_K^2)\right)\;.
 \label{8}
\end{eqnarray}
Keeping only up to quadratic corrections in $q$,
\begin{equation}
   P_1({\vec p}_1)
   \simeq \left[ 1 + {1\over 2} \left( q_\perp
                   + {q_s^2 \over 4 K_\perp}\right) {d\over dK_\perp}
                   + {q_L\over 2} {d \over dK_L}
                   + {1\over 2} \left( {q_\perp\over 2} {d \over dK_\perp}
                   + {q_L\over 2} {d\over dK_L} \right)^2 \right]\,
   P_1({\vec K})\, .
\label{9}
\end{equation}
$P_1({\vec p}_2)$ can be found simply by letting ${\vec q}\rightarrow
-{\vec q}$ in the above expression. Combining these,
re-exponentiating, and again keeping only terms up to second order in
$q_i$, we get
\begin{equation}
   P_1({\vec p}_1)\, P_1({\vec p}_2) \simeq [P_1({\vec K})]^2\,
   \exp\left[ - q_s^2\, \delta\!R_s^2 - q_\perp^2\, \delta\!R_\perp^2
      - q_L^2\, \delta\!R_L^2 - 2 q_\perp q_L\, \delta\!R_{\perp L}^2
             \right]\, ,
 \label{17}
\end{equation}
where
 \begin{eqnarray}
   \delta\!R_s^2
   & = &
   - {1 \over 4 K_\perp}\,\frac{d}{dK_\perp}
   \ln P_1({\vec K})\Bigl|_{K_L}\, ;
 \nonumber \\
   \delta\!R_\perp^2
   & = &
   - {1\over 4} \frac{d^2}{dK_\perp^2}
   \ln P_1({\vec K})\Bigl|_{K_L} \, ;
 \nonumber \\
   \delta\!R_L^2
   & = &
   - {1\over 4} \frac{d^2}{dK_L^2}
   \ln P_1({\vec K})\Bigr|_{K_\perp} \, ;
 \nonumber \\
   \delta\!R_{\perp L}^2
   & = &
   - {1\over 4} \frac{d}{dK_L}\,\left\{
     \frac{d}{dK_\perp} \ln P_1({\vec K})
     \Bigl|_{K_L}\right\}\Biggr|_{K_\perp}
   \, .
\label{17a}
\end{eqnarray}
Notice that all of these corrections are direct experimental
observables. For example, $\delta\!R_\perp^2$ is the curvature of a
plot of $\ln P_1({\vec p}\,)$ as a function of $p_\perp$ for fixed
$p_L$.

Finally, we turn to the corrections induced by using the correct
off-shell energy ${\textstyle\frac{1}{2}}(p_1^0+p_2^0)$ in the
emission function of eqn. (\ref{2}) rather than the approximate
on-shell value $E_K$ of eqn. (\ref{6}). Again making a Taylor
expansion for $|{\vec q}\,|\ll E_K$,
\begin{eqnarray}
{\textstyle\frac{1}{2}}(p_1^0+p_2^0) &\simeq&
E_K \left[ 1 + {1 \over 8E_K^2}
       \left(|{\vec q}\,|^2 -({\vec \beta}{\cdot}{\vec q}\,)^2 \right)
       \right]
\nonumber \\
&\simeq&
\sqrt{m^2 + |{\vec K}|^2 + {\textstyle\frac{1}{4}}\left(|{\vec q}\,|^2
                - ({\vec \beta}{\cdot}{\vec q}\,)^2\right)} \; ,
\label{18}
\end{eqnarray}
we can see that
\begin{equation}
   [{\textstyle\frac{1}{2}}(p_1+p_2)]^2 \simeq
\left[m^2 +
   {\textstyle\frac{1}{4}} \left( |{\vec q}\,|^2 - ({\vec
           \beta}{\cdot}{\vec q}\,)^2 \right)\right] \, .
\label{18a}
\end{equation}
Therefore, we can expand around the on-shell momentum $K$ in the
following way:
 \begin{equation}
   S\left(x,{\textstyle\frac{1}{2}}(p_1+p_2)\right)
   \simeq \left\{1 +
     {\textstyle\frac{1}{4}} \left( |{\vec q}\,|^2
                    - ({\vec \beta}{\cdot}{\vec q}\,)^2\right)
     {d\over dm^2} \right\}\, S(x,K)\;.
 \label{18b}
\end{equation}
To quadratic order in ${\vec q}$, then
 \begin{eqnarray}
   &&\int d^4x\, S\left( x,{\textstyle\frac{1}{2}}(p_1+p_2)\right) \,
     e^{iq{\cdot}x} \simeq
 \nonumber\\
   &&\exp\left[ {\textstyle\frac{1}{4}} \left(|{\vec q}\,|^2
               -({\vec \beta}{\cdot}{\vec q}\,)^2\right)
               {d\over dm^2}\ln P_1({\vec K}) \right]
   \int d^4x\, S(x,K) \, e^{iq{\cdot}x} \, .
 \label{19}
\end{eqnarray}

Putting everything together, we find the following corrected
model-independent expressions for the correlation radii of
eqn. (\ref{6g}):
\begin{eqnarray}
  R_s^2
  & \simeq &
  \langle y^2\rangle
  + \left( \frac{1}{4 K_\perp}\frac{d}{dK_\perp}
  - {1\over 2} \frac{d}{dm^2} \right) \ln P_1({\vec K}) \, ;
 \nonumber \\
  R_\perp^2
  & \simeq &
  \langle(x-\beta_\perp t)^2\rangle
  -\langle x-\beta_\perp t\rangle^2
 \nonumber \\
  & + &
  \left( {1\over 4} \frac{d^2}{dK_\perp^2}
  - {1\over 2}(1-\beta_\perp^2) {d\over dm^2}\right) \ln P_1({\vec K})
  \, ;
 \nonumber \\
  R_L^2
  & \simeq &
  \langle(z-\beta_L t)^2\rangle - \langle z-\beta_L t\rangle^2
 \nonumber \\
  & + &
  \left( {1\over 4} {d^2 \over dK_L^2}
  - {1\over 2} (1-\beta_L^2) {d\over dm^2} \right)
  \ln P_1({\vec K}) \, ;
 \nonumber \\
  R_{\perp L}^2
  & \simeq &
  \langle(x-\beta_\perp t)(z-\beta_L t)\rangle
  - \langle x-\beta_\perp t\rangle \langle z-\beta_L t\rangle
 \nonumber \\
  & + &
  \left( {1\over 4} {d^2 \over dK_\perp dK_L}
  + {1\over 2} \beta_\perp \beta_L {d\over dm^2} \right)
  \ln P_1({\vec K}) \, .
 \label{21}
\end{eqnarray}
Note that LCMS radii can be found from the above expressions simply by
setting $\beta_L=0$.

The first thing to observe about the above radii is that cylindrical
symmetry alone does not cause $R_{\perp L}^2$ to vanish, so a $q_\perp
q_L$ cross term (as in (\ref{6g})) should be included in any
experimental fit to the data.  However, it is interesting to note that
for the case $K_\perp = 0$ ($\beta_\perp=0$), $S(x,K)$ is independent
of $\phi$, so $R_{\perp L}^2$ does vanish (see appendix).  Furthermore
for this case $R_s^2 = R_\perp^2$ as it must, since if $K_\perp=0$ it
is impossible to define a difference between the ``out'' and ``side''
directions.  This means that the $q_\perp q_L$ cross term (as well as
the difference between $R_\perp^2$ and $R_s^2$) will be most
noticeable for pairs with large $K_\perp$.  We would also like to
point out that the cross term vanishes for spherically symmetric
systems if one redefines the $\hat{z}$ direction in the direction of
${\vec K}$ \cite{csorgo}, since in this case $K_\perp =0$ by
definition.  For any collision experiment, however, it is best not to
make this redefinition, since only cylindrical symmetry about the beam
can be assumed.  It should also be noted that if future heavy ion
experiments are able to generate HBT correlation functions from a
single event, then cross terms involving $q_s q_\perp$ and $q_s q_L$
should be included in any fits as tests of the cylindrical symmetry of
the individual reaction under consideration.

Before going on, we would like to say a few words about the validity
of the approximation of eqn. (\ref{6}) and the size of the correction
terms.  Since the $\delta\! R_i^2$ of (\ref{17a}) can be measured from
single particle distributions, a model-independent experimental
estimate can be made as to the accuracy of the approximation of
(\ref{6}) by comparing those correction terms with the HBT radii found
by fitting correlation data with gaussians as in (\ref{6g}).  If the
former are much smaller than the latter, then (\ref{6}) should be a
good approximation.  For example, the slopes and curvatures seen in
heavy ion collision data generate $\delta\! R_i^2$ which typically
have scales on the order of
\begin{equation}
|\delta\! R_i^2|\, {\mathrel{\lower.9ex\hbox{$\stackrel{\displaystyle
<}{\sim}$}}}\,\frac{1}{4(150 {\rm MeV})^2}\;,
\label{21a}
\end{equation}
whereas $R_s^2$, $R_\perp^2$ and $R_L^2$ typically have scales on the
order of $R_i^2\sim 1/(75 {\rm MeV})^2$ \cite{na35} so the
approximations to these radii from eq. (\ref{6}) should be good to
within roughly 5\%.  As we will see later, however, the corrections
could become important when determining the magnitude (and sign) of
the cross term or the difference between $R_\perp^2$ and $R_s^2$ for
systems with very short emission times.

\subsection{Boost Invariant Coordinates}
Now we would like to rederive the results of the preceding section
using rapidities rather than longitudinal momenta, since the former
boost invariant variables are usually more appropriate for
relativistic collision experiments.  Returning to eqn. (\ref{6}), let
us make an alternative on-shell definition of the 4-vector $K$:
\begin{equation}
K=(m_t\,{\rm ch}Y,\;{\vec K}_t,\;m_t\,{\rm sh}Y)
\label{2.2.1}
\end{equation}
where ${\vec K}_t={\textstyle\frac{1}{2}}({\vec p}_{1t}+{\vec
p}_{2t})$, $m_t^2=m^2+|{\vec K}_t|^2$, $Y={\textstyle\frac{1}{2}}({\rm
y}_1+{\rm y}_2)$, and ${\rm y}_i = {\textstyle\frac{1}{2}}\ln
[(E_i+p_{iL})/(E_i-p_{iL})]$.  Note that we use the subscript $t$
throughout to denote transverse 2-vectors as well as $m_t$ and other
general transverse quantities; this should not be confused with the
subscript $\perp$ which we use only to denote the ``out'' direction.

Just as in the last subsection, we can expand the factor
$\exp(iq{\cdot}x)$ in eqn. (\ref{6}) for small momentum (and rapidity)
differences.  This time we find:
\begin{eqnarray}
\tilde{C}({\rm y},q_s,q_\perp,Y,K_\perp) & \simeq &
1 \pm \Biggl\{1 - q_s^2\left\langle y^2\right\rangle
\\
&-&\left\langle \left[q_\perp\left(x-\frac{K_\perp}{m_t}\tau\,{\rm
ch}(\eta-Y)\right) +{\rm y}\,m_t\,\tau\,{\rm sh}(\eta-Y)
\right]^2\right\rangle
\nonumber \\
& + & \left\langle q_\perp\left(x-\frac{K_\perp}{m_t}\tau\,{\rm
ch}(\eta-Y)\right)
+{\rm y}\,m_t\,\tau\,{\rm sh}(\eta-Y)\right\rangle^2 \Biggr\}\;,
\nonumber
\label{2.2.2}
\end{eqnarray}
where ${\rm y}={\rm y}_1 - {\rm y}_2$, $\tau=\sqrt{t^2-z^2}$ is
longitudinal proper time, and
$\eta={\textstyle\frac{1}{2}}\ln[(t+z)/(t-z)]$ is the space-time
rapidity.  (The reader should take care not to confuse the rapidity
difference ${\rm y}$ with the cartesian coordinate $y$.)  This time
after exponentiating, we get a correlation function of the form:
\begin{equation}
\tilde{C}({\rm y},q_s,q_\perp,Y,K_\perp)
\simeq 1 \pm \exp\left[-q_s^2R_s^2 - q_\perp^2R_\perp^2
-{\rm y}^2\alpha^2 - 2q_\perp {\rm y} R_{\perp {\rm y}}\right] \;,
\label{2.2.3}
\end{equation}
where again for the approximation of eqn. (\ref{6}) the correlation
``radii'' can be read off as the coefficients of the appropriate terms
in eqn. (\ref{2.2.2}):
\begin{eqnarray}
R_s^2 & = &
\langle y^2\rangle
\nonumber \\
R_\perp^2 & = & \left\langle\left[x-(K_\perp/m_t)\tau{\rm
ch}(\eta-Y)\right]^2\right\rangle
\nonumber \\
& &-\left\langle x-(K_\perp/m_t)\tau{\rm ch}(\eta-Y)
\right\rangle^2
\nonumber \\
\alpha^2 & = & \left\langle\left[m_t\tau{\rm sh}(\eta-Y)\right]^2
\right\rangle
-\left\langle m_t\tau{\rm sh}(\eta-Y)\right\rangle^2
\nonumber \\
R_{\perp {\rm y}} & = & \left\langle\left[m_tx-K_\perp\tau{\rm ch}
(\eta-Y)
\right]\tau{\rm sh}(\eta-Y)\right\rangle
\nonumber \\
& &- \left\langle m_tx-K_\perp\tau{\rm ch}(\eta-Y)
\right\rangle
\left\langle\tau{\rm sh}(\eta-Y)\right\rangle\;.
\label{2.2.3a}
\end{eqnarray}

Similarly to eqn. (\ref{17}), quadratic corrections which arise from
expanding the denominator of (\ref{2}) for small $q$ can be found to
give
\begin{equation}
\frac{P_1({\vec p}_1)P_1({\vec p}_2)}{|\int d^4x S(x,K)|^2}\simeq
\exp\left(-q_s^2\delta\! R_s^2 -q_\perp^2\delta\! R_\perp^2
- {\rm y}^2 \delta \alpha^2
-2q_\perp {\rm y} \delta\! R_{\perp {\rm y}}\right)
\label{2.2.4}
\end{equation}
where
\begin{eqnarray}
\delta\! R_s^2 & = & -\frac{1}{4} \frac{1}{K_\perp}\,\frac{d}{dK_\perp}
\ln P_1(K)\Bigl|_Y
\nonumber \\
\delta\! R_\perp^2 & = & -\frac{1}{4} \frac{d^2}{dK_\perp^2}
\ln P_1(K)\Bigl|_Y
\nonumber \\
\delta \alpha^2 & = & -\frac{1}{4} \frac{d^2}{dY^2}
\ln P_1(K)\Bigr|_{K_\perp}
\nonumber \\
\delta\! R_{\perp {\rm y}} & = & -\frac{1}{4} \frac{d}{dY}\,\left\{
\frac{d}{dK_\perp}\ln P_1(K)\Bigl|_Y\right\}\Biggr|_{K_\perp}
\label{2.2.5}
\end{eqnarray}
Note that these ``side'' and ``out'' corrections take the same form as
those in eqn. (\ref{17a}), except that here rapidity rather than
longitudinal momentum is held fixed while taking the derivative with
respect to $K_\perp$.  Since experimental one particle spectra are
usually presented as functions of rapidity and not longitudinal
momentum, these corrections can be even more readily measured from the
data than those of the previous subsection.

Finally, we turn again to the corrections induced by using the exact
off-shell 4-vector ${\textstyle\frac{1}{2}}(p_1+p_2)$ in the emission
function of eqn. (\ref{2}) rather than the approximate on-shell
4-vector $K$ of eqn. (\ref{2.2.1}).  Making a Taylor expansion for
small ${\rm y}$ and ${\vec q}_t$, we find
\begin{equation}
\left[{\textstyle\frac{1}{2}}(p_1+p_2)\right]^2 \simeq m^2
+ {\textstyle\frac{1}{4}} q_\perp^2\left(1-\frac{K_\perp^2}{m_t^2}
\right)
+{\textstyle\frac{1}{4}} q_s^2 + {\textstyle\frac{1}{4}}
{\rm y}^2\,m_t^2 \;.
\label{2.2.6}
\end{equation}
Furthermore, if we reparametrize the local 4-vectors of (\ref{6b}) in
the following way
\begin{equation}
u(x) = \left(u_t\,{\rm ch}\xi,u_\rho \cos\phi, u_\rho \sin\phi,
u_t\,{\rm sh}\xi\right)
\label{2.2.7}
\end{equation}
where $u_t$, $u_\rho$ and $\xi$ are independent of $\phi$, then
\begin{equation}
{\textstyle\frac{1}{2}} (p_1+p_2){\cdot}u \simeq \left[ m_t^2
+{\textstyle\frac{1}{4}}
q_\perp^2\left(1-\frac{K_\perp^2}{m_t^2}\right)
+{\textstyle\frac{1}{4}} q_s^2 + {\textstyle\frac{1}{4}} {\rm
y}^2\,m_t^2\right]^{1/2}\!\!  u_t\,{\rm ch}(Y-\xi) -K_\perp
u_\rho\cos\phi\;.
\label{2.2.8}
\end{equation}
Therefore, to quadratic order in ${\rm y}$ and ${\vec q}_t$
\begin{equation}
\frac{\int d^4x S[x,{\textstyle\frac{1}{2}}(p_1+p_2)]
e^{iq{\cdot}x}}{\int d^4x
S(x,K)e^{iq{\cdot}x} } = \exp\left\{
{\textstyle\frac{1}{4}}\left[q_\perp^2\left(1-\frac{K_\perp^2}
{m_t^2}\right)
+ q_s^2 + {\rm y}^2\,m_t^2\right]\frac{d}{dm^2}\ln P_1(K)\right\} \;.
\label{2.2.9}
\end{equation}
The most interesting thing to note about this off-shell correction is
that it has no effect on the coefficient of the $q_\perp {\rm y}$ cross
term.

Putting everything together, we find the following corrected
model-independent expressions for the correlation radii of
eqn. (\ref{2.2.4}):
\begin{eqnarray}
R_s^2 & \simeq &
\langle y^2\rangle +
\,\left(\frac{1}{4}\frac{1}{K_\perp}\frac{d}{dK_\perp}
-\frac{1}{2}\frac{d}{dm^2}\right)\ln P_1({\vec K})
\nonumber \\
R_\perp^2 & \simeq & \left\langle\left[x-\frac{K_\perp}{m_t}\tau{\rm
ch}(\eta-Y)\right]^2\right\rangle
-\left\langle x-\frac{K_\perp}{m_t}\tau{\rm ch}(\eta-Y)
\right\rangle^2
\nonumber \\
& + & \left(\frac{1}{4}\frac{d^2}{dK_\perp^2}
-\frac{1}{2}(1-\frac{K_\perp^2}{m_t^2})
\frac{d}{dm^2}\right)\ln P_1({\vec K})
\nonumber \\
\alpha^2 & \simeq & \left\langle
\left[m_t\tau{\rm sh}(\eta-Y)\right]^2\right\rangle
-\left\langle m_t\tau{\rm sh}(\eta-Y)\right\rangle^2
\nonumber \\
& + & \left(\frac{1}{4}\frac{d^2}{dY^2}
-\frac{1}{2} m_t^2\frac{d}{dm^2}\right)\ln P_1({\vec K})
\nonumber \\
\vspace*{.2in}
R_{\perp {\rm y}} & \simeq &
\Bigl\langle\left[m_t\,x -K_\perp\tau{\rm ch}(\eta-Y)
\right]\tau{\rm sh}(\eta-Y)\Bigr\rangle
\nonumber \\
&-& \left\langle m_t\,x -K_\perp\tau{\rm ch}(\eta-Y)
\right\rangle
\left\langle\tau{\rm sh}(\eta-Y)\right\rangle
+ \frac{1}{4}\frac{d}{dY} \frac{d}{dK_\perp}\ln P_1({\vec K})
\label{2.2.10}
\end{eqnarray}
Again, although $R_{\perp {\rm y}}$ does not vanish in general, it does
vanish for pairs with $K_\perp = 0$ (see appendix).

\section{A Model with Nonrelativistic Expansion}

To get an idea of the usefulness of the model independent expressions
just derived, we study a slight generalization of the thermal
emission function presented in \cite{csorgo}:
\begin{equation}
   S(x,K) = \frac{E_K}{(2\pi)^3}
   \exp\left(-\frac{K{\cdot}u(x)}{T}\right)\,
   H(t)\, I(\rho)\, J(z) \, .
 \label{2.2}
\end{equation}
Here $T$ is a constant freeze-out temperature, and we define the
space-time distribution of the source by a product of gaussians in the
center of mass frame of an expanding fireball
\begin{equation}
   H(t)I(\rho)J(z) =
   \frac{1}{\sqrt{2\pi(\delta t)^2}}
   \exp\left(-\frac{(t-t_0)^2}{2(\delta t)^2}-\frac{\rho^2}{2R_G^2}
             -\frac{z^2}{2L_G^2}\right)\,.
 \label{2.3}
\end{equation}
For the thermal smearing factor in eqn.~(\ref{2.2}), we take a
nonrelativistic linear expansion 4-velocity
\begin{eqnarray}
u(x)&=&\left[1-(v_R\,\rho/R_G)^2-(v_L\,z/L_G)^2\right]^{-1/2}\,
  \left(1,\, v_R\,x/R_G,\, v_R\,y/R_G,\,v_L\,z/L_G \right)\;,
\nonumber \\
&\simeq& \left(
   1+{\textstyle\frac{1}{2}}(v_R\,\rho/R_G)^2+
{\textstyle\frac{1}{2}}(v_L\,z/L_G)^2,\,
   v_R\,x/R_G,\, v_R\,y/R_G,\,v_L\,z/L_G \right)\;, \label{2.1}
\end{eqnarray}
where $v_R\ll 1$ and $v_L\ll 1$ are the transverse and longitudinal
flow velocities of the fluid at $\rho=R_G$ and $z=L_G$, respectively.

Note that in the limit $\delta t \rightarrow 0$, $S(x,K)$ becomes the
Boltzmann approximation to the hydrodynamic emission function of
eqn.~(\ref{4}) with a constant freeze-out time $t_0$ and a local
chemical potential given by:
\begin{equation}
   \frac{\mu(x)}{T} = -\frac{\rho^2}{2R_G^2} - \frac{z^2}{2L_G^2}\, .
 \label{2.3aa}
\end{equation}
In a sense, use of a nonzero $\delta t$ can be thought of as a
smearing of the sharp 3-dimensional freeze-out hypersurface $t=t_0$
over the fourth (temporal) dimension.

Since the model is completely gaussian, analytic calculation of the
one-particle distribution is straightforward, yielding
\begin{equation}
   P_1({\vec K}) = \frac{E_K}{(2\pi)^{3/2}}\,
   R_*^2\, L_*\,
   \exp\left(-\frac{E_K}{T} + \frac{R_*^2v_R^2K_\perp^2}{2R_G^2T^2}
    + \frac{L_*^2v_L^2K_L^2}{2L_G^2T^2}\right)\, ,
 \label{2.4}
\end{equation}
where
\begin{equation}
   \frac{1}{R_*^2} = \frac{1}{R_G^2}\left(1 + \frac{E_K}{T}v_R^2
\right)\, ;
   \qquad
   \frac{1}{L_*^2} = \frac{1}{L_G^2}\left(1 + \frac{E_K}{T}v_L^2
\right)\, .
 \label{2.5}
\end{equation}

\begin{figure}
\centerline{
\epsfxsize=4.5in
\epsfbox{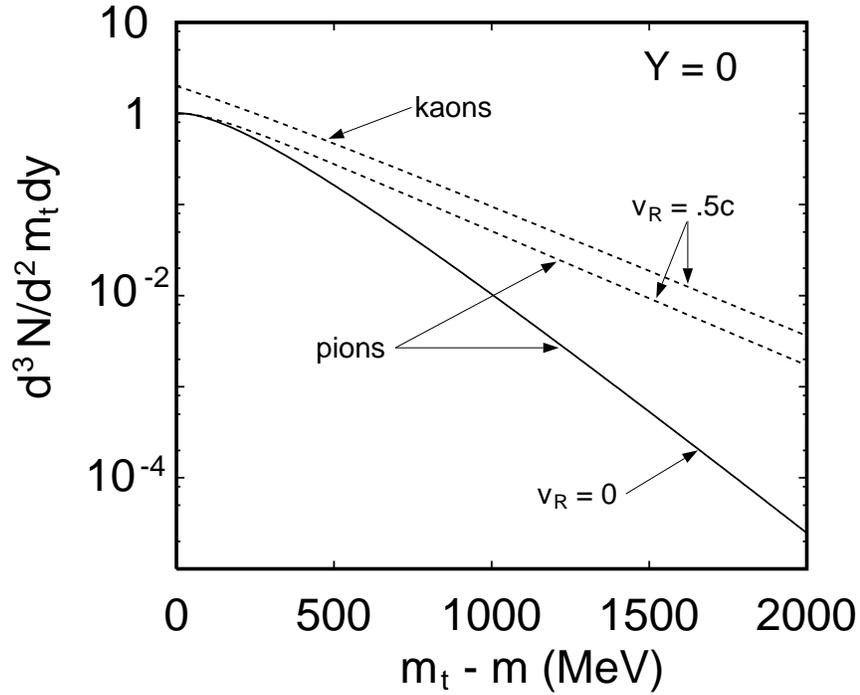}
}
\caption{The one particle spectrum of (47) is plotted as a
function of $m_t-m$ for midrapidity ($Y=0$) pions and kaons.  The
solid curve is for pions with no transverse flow ($v_R=0$), while the
dashed curves are for pions (intercept normalized to 1) and kaons
(intercept normalized to 2) with $v_R=0.5c$.  The other source
parameters used are $R_G=L_G=3$ fm, $T=150$ MeV, and $v_L=0$.}
\end{figure}

\begin{figure}
\centerline{
\epsfxsize=4.5in
\epsfbox{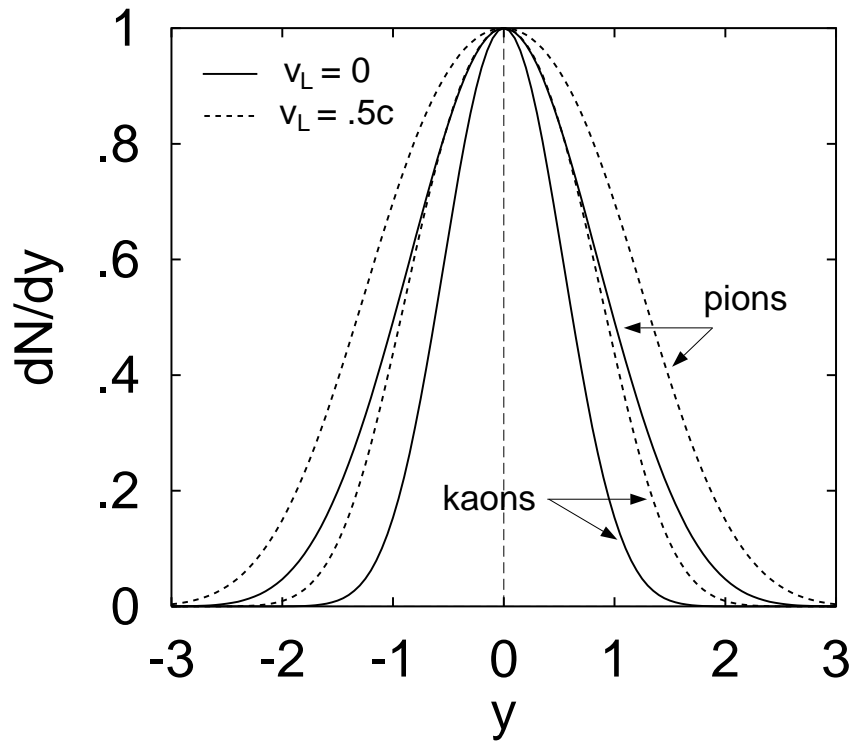}
}
\caption{The effect of longitudinal flow on the rapidity
distribution of pions (outer two curves) and kaons (inner two curves)
is shown for a source with no transverse flow $v_R=0$.  The solid
curves are for $v_L=0$ while the dashed curves are for $v=0.5c$.  The
other source parameters are defined as in fig. 1, and all curves have
been normalized to 1 at $Y=0$.}
\end{figure}

In fig. 1 we plot $P_1({\vec K})$ as a function of $m_t-m$ for
midrapidity ($Y=0$) particles from a source with the parameters
$R_G=L_G=3$ fm, $T=150$ MeV, and $v_L=0$.  The decrease in the slope
of the pion curves when the transverse flow is changed from $v_R=0$ to
$v_R=0.5c$ can be understood in terms of an effective blueshifted
temperature
\cite{schnekki}
\begin{equation}
T_{\rm eff} = T\sqrt{\frac{1+v_R}{1-v_R}}\;.
\label{2.5a}
\end{equation}
{}From eq. (\ref{2.5}) it can also be seen that asymptotically as
$K_\perp\rightarrow \infty$
\begin{equation}
R_*^2 \rightarrow \frac{R_G^2 T}{K_\perp v_R^2}
\label{2.5b}
\end{equation}
so that the $K_\perp$ dependence of the prefactor drops out and the
spectrum takes the form of a pure exponential with an inverse slope of
$T_{\rm eff}=2T$.  Figure 1 also features a kaon distribution with
flow which shows the same behavior.  As can be seen from fig. 2,
increasing the amount of longitudinal flow from $v_L=0$ to $v_L=0.5c$
causes a widening of the rapidity distribution both for pions and
kaons.

Using (\ref{21}) the correlation radii are readily found to be:
\begin{eqnarray}
   R_s^2
   & = &
   R_*^2 + \left[\frac{R_*^2v_R^2}{4R_G^2T^2} \right]\, ;
 \nonumber \\
   R_\perp^2
   & = &
   R_*^2 + \beta_\perp^2(\delta t)^2
   - \left[\frac{\beta_\perp^2}{4E_K^2} - \frac{R_*^2v_R^2}{4R_G^2T^2}
   + {\cal O}\left(v_R^4,v_L^4\right)\right]\, ;
 \nonumber \\
   R_L^2
   & = &
   L_*^2 + \beta_L^2(\delta t)^2
   - \left[\frac{\beta_L^2}{4E_K^2} - \frac{L_*^2v_L^2}{4L_G^2T^2}
   + {\cal O}\left(v_R^4,v_L^4\right)\right]\, ;
 \nonumber \\
   R_{\perp L}^2
   & = &
   \beta_\perp \beta_L(\delta t)^2
   - \left[\frac{\beta_\perp\beta_L}{4E_K^2}
   + {\cal O}\left(v_R^2+v_L^2\right)^2\right]\, ;
\label{2.6}
\end{eqnarray}
where the corrections due to using eqn.~(\ref{2}) rather than
eqn.~(\ref{6}) have been grouped in the square brackets, and we have
only kept terms up to second order in the velocities in order to be
consistent with the nonrelativistic approximation (\ref{2.1}).

First we would like to note that by rotating the coordinate system for
each pair to $\hat{z}=\hat{K}$ (so that $K_\perp=0$ by definition) and
neglecting the correction terms, our expressions for the correlation
radii reduce to those of \cite{csorgo} for the nonrelativistic
($E_K=m$) and spherically symmetric case of $R_G=L_G$ and
$v_R/R_G=v_L/L_G=1/t_0$.  As they point out and can be seen from eq.
(\ref{2.5}), transverse flow causes the ``side'' radius to measure
something smaller than the real geometrical radius $R_G$.  In fact
since this is a completely gaussian model, it should be no surprise
that $R_s$ simply measures the length of homogeneity of eq.
(\ref{7c}) in the transverse direction.  Similarly, $L_*^2$ just
measures the longitudinal length of homogeneity which will be smaller
than $L_G$ if $v_L>0$.  One of the most interesting features of
(\ref{2.6}) is that in the absence of the corrections, not only is the
difference between $R_\perp^2$ and $R_s^2$ directly proportional to
the square of the emission time (which is simply the length of
homogeneity in the temporal direction), $R_{\perp L}^2$ is as well.

Notice that for systems with $T \sim 150$ MeV and $R_s,\;R_\perp,\;R_L
\sim 3$ fm, the correction terms do not alter the naive expressions
for those radii by more than 3\%.  For very small radii and very short
emission times $\delta t$, however, the correction terms may actually
have a noticeable cancellation effect both on the magnitude of the
cross term and on the difference between $R_\perp^2$ and $R_s^2$.
This should be kept in mind when extracting limits on $\delta t$ from
the data
\cite{na35}.  For example,
for pions with $K_\perp\sim m$, this kind of cancellation will occur
for emission times $\delta t<{\textstyle\frac{1}{2}}$ fm.  In
particular, for $\delta t=0$, $R_\perp^2$ would actually be smaller
than $R_s^2$ in this model.  However, since present heavy ion
correlation radii are measured to be around 3 fm \cite{na35} and the
experiments are not yet able to resolve 3\% effects, keeping the
correction terms may not be necessary when comparing a specific model
to heavy ion correlation data.

One might at first think that the cross term for this model would
vanish if the radii were calculated in the LCMS frame, since
$\beta_L=0$ in that frame.  This is not the case, however, because the
emission function $S(x,K)$ is not longitudinally boost invariant, even in the
case of non-relativistic Galilei-transformations. After making
the appropriate transformations into the LCMS frame
\begin{equation}
t^\prime = \gamma_L(t-\beta_L z)\;\;\;\;\;\;\;\;\;
z^\prime =\gamma_L(z-\beta_L t)\;\;\;\;\;\;\;\;\;
\gamma_L = (1-\beta_L^2)^{-1/2}\;,
\label{2.7}
\end{equation}
$t^\prime\,z^\prime$ cross terms are introduced into the gaussians.
These in turn give rise not only to a nonzero $R_{\perp L}^2$ cross term
but also modifications to the other radii.  Neglecting the correction
terms,
\begin{eqnarray}
   R_s^{\prime 2}
   & = &
   R_*^2
 \nonumber \\
   R_\perp^{\prime 2}
   & = &
   R_*^2 + \beta_\perp^2\gamma_L^2
   \left[(\delta t)^2+\beta_L^2L_*^2\right]
 \nonumber \\
   R_L^{\prime 2}
   & = &
   \gamma_L^2\left[L_*^2 + \beta_L^2(\delta t)^2\right]
 \nonumber \\
   R_{\perp L}^{\prime 2}
   & = &
   \beta_\perp \beta_L\gamma_L^2\left[(\delta t)^2 + L_*^2\right]
\label{2.8}
\end{eqnarray}
where $\beta_L$ and $\gamma_L$ in the above expressions are evaluated
in the fixed center of mass frame.  Note that in this frame there is
now also a geometrical contribution $\sim L_*^2$ to $R_{\perp
L}^{\prime 2}$ and $R_\perp^{\prime 2}-R_s^{\prime 2}$; since it is multiplied
by a factor $1/c^{2}$ relative to the $(\delta t)^2$ terms, it vanishes in the
non-relativistic limit $c\to\infty$. However, the $(\delta t)^2$ contribution
to $R_{\perp L}^{\prime 2}$ in particular survives in this limit.

\section{A Model with Relativistic Longitudinal Expansion}

Now we move to a model similar to those in \cite{csorgo2} which should
provide a more realistic description of particle emission from a
relativistic collision.  In the center of mass frame of an expanding
fireball, we define the following emission function
\begin{equation}
S(x,K) = \frac{\tau_0\,m_t\,{\rm
ch}(\eta-Y)}{(2\pi)^3\,\tau\,\sqrt{2\pi(\delta \tau)^2}}
\exp\left[-\frac{K{\cdot}u(x)}{T}\right]
\exp\left[-\frac{(\tau-\tau_0)^2}{2(\delta \tau)^2}
-\frac{\rho^2}{2R_G^2}-\frac{\eta^2}{2(\delta \eta)^2}
\right]\;,
\label{4.2}
\end{equation}
where again $T$ is a constant freeze-out temperature,
$\tau=\sqrt{t^2-z^2}$ is the longitudinal proper time, and
$Y={\textstyle\frac{1}{2}}\ln[(E_K+K_L)/(E_K-K_L)]$ is the rapidity of
a particle with momentum ${\vec K}$.  This time in the limit $\delta
\tau\rightarrow 0$, (\ref{4.2}) becomes the Boltzmann approximation
to (\ref{4}) with a constant freezeout {\em proper} time $\tau_0$ and
a local chemical potential given by
\begin{equation}
\frac{\mu(x)}{T} = -\frac{\rho^2}{2R_G^2} - \frac{\eta^2}{2(\delta\eta)^2}
\label{4.1a}
\end{equation}
The second exponential in the emission function (\ref{4.2}) can be interpreted
as the space-time distribution of point-like sources, each of which emits a
thermal spectrum, boosted by the flow 4-velocity $u(x)$, as given by the first
exponential and the ${\rm ch}(\eta-Y)$ prefactor. For simplicity, the source
distribution in space-time is taken to be gaussian.\\
For this model, we consider a flow which is still non-relativistic
transversally but which now exhibits Bjorken expansion (fluid rapidity
= space-time rapidity) longitudinally,
\begin{eqnarray}
u(x) \simeq \left(\left(1+{\textstyle\frac{1}{2}}(v\,\rho/R_G)^2
\right){\rm ch}\eta
,\,(v\,x/R_G),
(v\,y/R_G),\,
\left(1+{\textstyle\frac{1}{2}}(v\,\rho/R_G)^2\right)
{\rm sh}\eta\,\right)\;,
\label{4.2a}
\end{eqnarray}
where $v\ll 1$ is the transverse flow velocity of the
fluid at $\rho = R_G$. This flow profile corresponds to a longitudinal velocity
$v_{L}(z,t)=z\,/\,t$ . With this definition, $K{\cdot}u$ takes the
following longitudinally boost-invariant form
\begin{equation}
K{\cdot}u = m_t[1+{\textstyle\frac{1}{2}}(v\,\rho/R_G)^2]{\rm ch}(\eta-Y)
-K_\perp (v\,x/R_G)\;.
\label{4.3}
\end{equation}

If we restrict ourselves to particle pairs with
$m_t\,{\mathrel{\lower.9ex\hbox{$\stackrel{\displaystyle >}{\sim}$}}}\,T$
and $|Y|\ll 1+ (\delta\eta)^2m_t/T$, then we can perform a modified
saddle point approximation by expanding ${\rm ch}(\eta-Y)$ in
(\ref{4.3}) in powers of $\eta^\prime=\eta-Y$, keeping in the exponent
only terms up to second order and expanding everything else to the
desired order.  For our calculations, we approximate $S(x,K)$ by

\begin{eqnarray}
S(x,K) &\simeq& \frac{m_t\tau_0(1+{\textstyle\frac{1}{2}}\eta^{\prime 2})}
{(2\pi)^3\,\tau\,\sqrt{2\pi (\delta\tau)^2}}
\exp\left[-\frac{m_t}{T}\left(1+ \frac{(v\rho)^2}{2R_G^2}\right)
\left(1+{\textstyle\frac{1}{2}} \eta^{\prime 2}\right)
+ \frac{K_\perp\,v\,x}{R_GT}\right]
\nonumber \\
&\times&
\left(1-\frac{m_t}{24T}\eta^{\prime 4}\right)
\exp\left[-\frac{(\tau-\tau_0)^2}{2(\delta \tau)^2}
-\frac{\rho^2}{2R_G^2}-\frac{(\eta^\prime +Y)^2}
{2(\delta \eta)^2}\right]\;.
\label{4.4}
\end{eqnarray}
Note that we keep only the $\eta^{\prime 2}$ term when expanding the
${\rm ch}\eta^\prime$ prefactor, but we also keep a term
$(m_t/T)\eta^{\prime 4}$ from the expansion of the exponent.  The
latter term is to be taken as be roughly of the same order as
$\eta^{\prime 2}$ for reasons which will become clear later.

For the particle emission time, it must be true physically that
$\delta\tau/\tau_0< 1$.  Rather than demanding the much stricter
condition $\delta\tau/\tau_0\ll 1$, we simply assume that this ratio
is small enough (e.g.  $\delta\tau/\tau_0\,
{\mathrel{\lower.9ex\hbox{$\stackrel{\displaystyle <}{\sim}$}}}
\,{\textstyle\frac{1}{2}}$) so that
we can replace integrals over only positive values of $\tau$ with ones
ranging from $-\infty$ to $+\infty$.  Finally, in all of our
calculations we throw away all terms of ${\cal O}(v^4)$, in keeping
with our nonrelativistic approximation in the transverse direction.

Given these approximations, calculation of the one particle
distribution can now be done analytically, yielding
\begin{eqnarray}
P_1(K) &\simeq& \frac{\tau_0 m_t}{(2\pi)^{3/2}}R_*^2(\delta\eta)_*
\left(1 + {\textstyle\frac{1}{2}}\frac{R_*^2}{R_G^2}
(\delta\eta)_*^2
-\frac{m_t}{8T}(\delta\eta)_*^4
\right)
\nonumber \\
& &\times\exp\left[-\frac{m_t}{T}+\frac{K_\perp^2(R_*v)^2}{2(R_GT)^2}
-\frac{Y^2}{2(\delta\eta)^2}\left(1-\frac{(\delta\eta)_*^2}
{(\delta\eta)^2}\right)\right]
+ {\cal O}\left[(\delta\eta)_*^5\right]
\label{4.5}
\end{eqnarray}
where
\begin{equation}
\frac{1}{R_*^2} = \frac{1}{R_G^2}\left(1 + \frac{m_t}{T}v^2\right)
\label{4.6}
\end{equation}
and our expansion parameter is defined by
\begin{equation}
\frac{1}{(\delta\eta)_*^2} = \frac{1}{(\delta\eta)^2} + \frac{m_t}{T}
\label{4.7}
\end{equation}
Note that for pairs in which $m_t/T \gg 1/(\delta\eta)^2$ as were
studied in \cite{sinyu2}, $(\delta\eta)_*^2$ becomes simply $T/m_t$.
This is the reason that we consider $(m_t/T)(\delta\eta)_*^4$ to be of
the same order as $(\delta\eta)_*^2$.

\begin{figure}
\centerline{
\epsfxsize=4.5in
\epsfbox{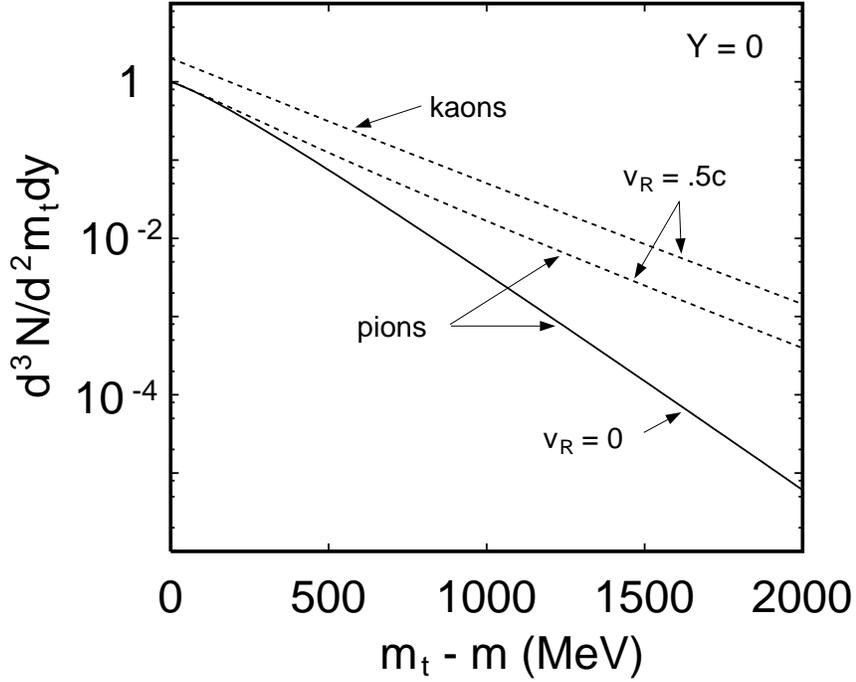}
}
\caption{The one particle spectrum obtained by numerically
integrating (54) is plotted as a function of $m_t-m$ for midrapidity
($Y=0$) pions and kaons.  The solid curve is for pions with no
transverse flow ($v=0$), while the dashed curves are for pions
(normalized to 1) and kaons (normalized to 2) with $v=0.5c$.  The other
source parameters used are $\tau_0=4$ fm/c, $R_G=3$ fm,
$\delta\eta=1.5$, and $T=150$ MeV.}
\end{figure}

\begin{figure}
\centerline{
\epsfxsize=4.5in
\epsfbox{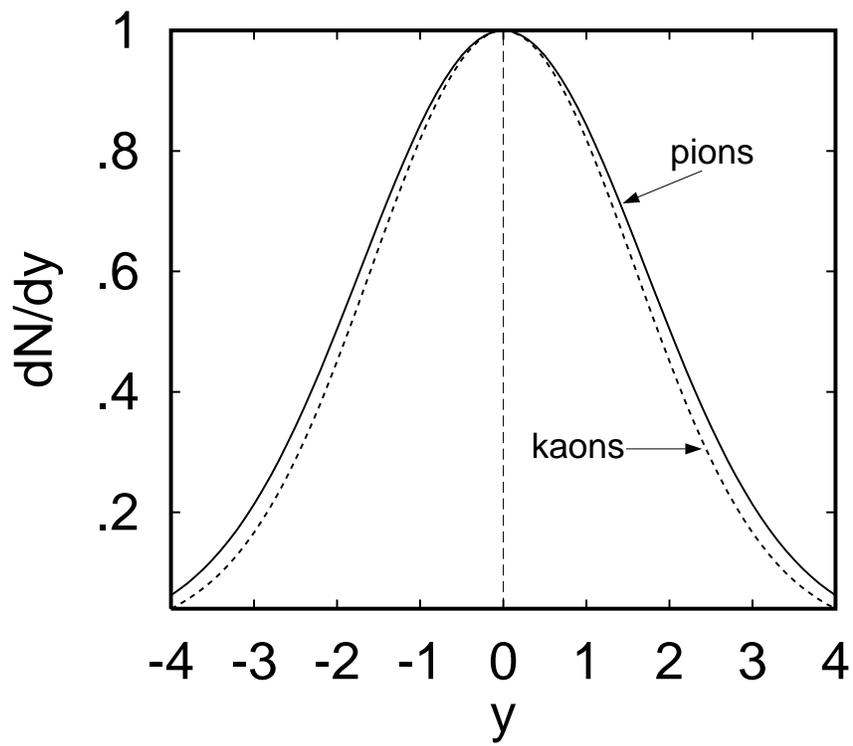}
}
\caption{The rapidity distribution of pions (solid curves)
and kaons (dashed curves) is shown for a source with no transverse
flow $v=0$ and other source parameters equal to those of fig. 3.}
\end{figure}

In fig. 3 we plot numerical calculations of $P_1({\vec K})$ as a
function of $m_t-m$ for midrapidity ($Y=0$) particles from a source
(\ref{4.2}) with the parameters $\tau_0=4$ fm/c, $R_G=3$ fm,
$\delta\eta=1.5$, and $T=150$ MeV.  We have checked that our analytic
expressions provide excellent ($< 5\%$ error) approximations to the
exact numerical results.  Again the decrease of the pion slope as the
transverse flow parameter $v$ is increased from $v=0$ to $v=0.5c$ can
be well understood in terms of the effective blueshifted temperature
of eq. (\ref{2.5a}).  In contrast to the nonrelativistic case,
however, $P_1$ does not quite become a pure exponential as
$K_\perp\rightarrow \infty$, because although the prefactor $R_*^2$
again cancels the prefactor $m_t$, there is an additional $K_\perp$
dependence coming from the prefactor
\begin{equation}
(\delta\eta)_* \rightarrow \sqrt{T/m_t}\;.
\label{4.7a}
\end{equation}
The effect of this prefactor can also be seen at the origin, where its
partial cancellation of the $m_t$ prefactor results in less curvature
than is seen in the nonrelativistic case of fig. 1.  For completeness,
in fig. 4 we also show the rapidity spectrum both for pions and kaons.
Although there is a slight decrease in the width for kaons, the effect
is much smaller in this case than in fig. 2 due to the relativistic
longitudinal flow which causes the difference in mass to become less
important.

The correlation radii can now be calculated by using
eqs. (\ref{21}) or (\ref{2.2.10}). By means of a saddle point approximation,
the correlation radii are expanded to the second order in the small parameters
$(\delta\eta)_*^2$ and $(\delta\tau/\tau_{0})^2$. Therefore, in performing
these calculations
below, we only keep terms of order $(\delta\eta)_*^4$ (or
$(m_t/T)(\delta\eta)_*^6$) except for smaller terms involving
$(\delta\tau/\tau_0)^2$ or $v^2$, in which case we keep only terms
of order $(\delta\eta)_*^2$ and $(\delta\eta)_*^0$, respectively. Please note
that, since the general expressions (\ref{21}) or (\ref{2.2.10}) for the
correlation radii are all at most quadratic in the proper time,
no term involving $(\delta\tau/\tau_0)^4$ will in fact occur in the results.

\subsection{HBT Radii in Cartesian Coordinates}
In cartesian coordinates, the correlation radii take the following form
(ordered by powers of the small expansion parameters $(\delta\tau/\tau_0)^2$
and $(\delta\eta)_*^2$):

\begin{eqnarray}
R_s^2 & = & R_*^2\;;
\nonumber \\
R_\perp^2 & = & R_*^2 + \frac{K_\perp^2}{m_t^2}
(\delta \tau)^2 + \frac{K_\perp^2}{m_t^2}\beta_L^2
\tau_0^2 (\delta\eta)_*^2 \nonumber\\
&+ &\frac{K_\perp^2}{m_t^2} \left(1+\beta_L^2
-2\beta_L\frac{Y}{(\delta\eta)^2}\right)(\delta \tau)^2(\delta\eta)_*^2
\nonumber \\
&+&\frac{K_\perp^2}{m_t^2}\tau_0^2\left[
\beta_L^2 \nu
-2\beta_L\frac{Y}{(\delta\eta)^2}
+{\textstyle\frac{1}{2}}\right](\delta\eta)_*^4\;;
\nonumber \\
R_L^2 & = & \frac{m_t^2}{E_K^2}
\tau_0^2 (\delta\eta)_*^2 + \frac{m_t^2}{E_K^2}(\delta\tau)^2
(\delta\eta)_*^2+
\frac{m_t^2}{E_K^2}
\nu \tau_0^2 (\delta\eta)_*^4\;;
\nonumber \\
R_{\perp L}^2 & = &
-\beta_\perp\beta_L\tau_0^2 (\delta\eta)_*^2
-\beta_\perp
\left[\beta_L -\frac{Y}{(\delta\eta)^2}\right](\delta\tau)^2(\delta\eta)_*^2
\nonumber \\
&-&\beta_\perp\tau_0^2
\left[\beta_L \nu-\frac{Y}{(\delta\eta)^2}\right](\delta\eta)_*^4\;.
\label{4.8}
\end{eqnarray}
Here $\nu
=1+(R_*/R_G)^2-{\textstyle\frac{1}{2}}(m_t/T)(\delta\eta)_*^2$, and we
have neglected the corrections which come from using (\ref{2}) instead
of (\ref{6}).  Although this model is not completely gaussian, within
the scope of our approximation $R_s^2$ still roughly measures the
transverse region of homogeneity of the fluid, as can be seen by
comparing (\ref{4.6}) with (\ref{7c}). Although we will show that in practice
all terms given in (\ref{4.8}) are important, we will for didactical purposes
first consider only the leading order in the small expansion parameters. Then
the expressions (\ref{4.8}) simplify and can be reformulated as follows:
\begin{eqnarray}
\frac{1}{R_{s}^2}&=&\frac{1}{R_{\perp}^2}\;\; =\;\;
\frac{1}{R_{G}^2}+\frac{m_{t}}{T}\frac{v^2}{R_{G}^2}
\nonumber\\
\frac{1}{R_{L}^2}&=& {\rm ch}^2 Y \left( \frac{1}{\tau_{0}^2(\delta
\eta)^2}+\frac{m_{t}}{T}\frac{1}{\tau_{0}^2}\right)
\nonumber\\
\frac{1}{R_{\perp L}^2}&=& -\frac{m_{t}}{K_{\perp}}\frac{{\rm ch}^2 Y}{{\rm sh}
Y}\left( \frac{1}{\tau_{0}^2(\delta
\eta)^2}+\frac{m_{t}}{T}\frac{1}{\tau_{0}^2}\right)
\label{ann}
\end{eqnarray}
In agreement with \cite{csorgo2} we find that, in each
principal direction of the expanding fireball, two length scales should be
distinguished. In addition to the geometric length
scales $R_{G}$ and $L_{G} = \tau_{0}\;\delta\eta\;$ in the transverse and
longitudinal directions respectively, we have two
``lengths of homogeneity" generated by the flow gradients. The transversal and
longitudinal homogeneity lengths are given by the following expressions:
\begin{eqnarray}
R_{H}^2=\frac{T}{m_{t}}\frac{R_{G}^2}{v^2}\;\;\;\;,\;\;\;\;
L_{H}^2=\frac{T}{m_{t}}\tau_{0}^2\;\;\;.
\label{homog}
\end{eqnarray}
Please note that the occurrence of $\tau_{0}$ in both longitudinal lengths,
$L_{G}$ and $L_{H}$, has two different origins: whereas the geometrical
longitudinal extension of the fireball at freeze-out is clearly always
proportional to the mean freeze-out proper time $\tau_{0}$, its occurrence in
the longitudinal homogeneity length is due to the specific choice of the
velocity profile, since for a longitudinally boost invariant velocity profile
the velocity gradient is just given by the inverse proper time.
In fact, the true origin of the homogeneity lengths (\ref{homog}) is seen by
writing them in the form
\begin{eqnarray}
R_{H}^2&=&\frac{T}{m_{t}}\frac{1}{(\partial v_t/\partial\rho)^2}\;,\nonumber\\
L_{H}^2&=&\frac{T}{m_{t}}\frac{1}{(\partial_{\mu}u_{L}^{\mu})^2}
\Biggr|_{\tau=\tau_{0}}\;,
\label{homogg}
\end{eqnarray}
where $v_t(\rho)=v\,\rho/R_G\,$, and in the second line $u_{L}^{\mu}=({\rm
ch}\eta, 0,0,{\rm sh}\eta)$ denotes the longitudinal part of the flow velocity
profile (\ref{4.2a}) which satisfies $\partial_{\mu}u_{L}^{\mu}=1/\tau$. Eq.
(\ref{homogg}) makes the nature of the homogeneity lengths explicit in showing
how they are generated by the flow gradients at freeze-out.\\
With these notations, the correlation radii can be written as follows :
\begin{eqnarray}
\frac{1}{R_{s}^2}&=&\frac{1}{R_{G}^2}+\frac{1}{R_{H}^2}\;;\nonumber\\
\frac{1}{R_{\perp}^2}&=&\frac{1}{R_{G}^2}+\frac{1}{R_{H}^2}\;;\nonumber\\
\frac{1}{R_{L}^2}&=&{\rm ch}^2
Y\left(\frac{1}{L_{G}^2}+\frac{1}{L_{H}^2}\right)\;;\nonumber\\
\frac{1}{R_{\perp L}^2}&=&-\frac{m_{t}}{K_{\perp}}\frac{{\rm ch}^2 Y}{{\rm
sh}Y} \left(\frac{1}{L_{G}^2}+\frac{1}{L_{H}^2}\right)\;.
\label{rewr}
\end{eqnarray}
As already pointed out in
\cite{csorgo2}, the correlation radii are seen to be dominated by the shorter
of the geometric and homogeneity lengths.
This means in particular that if $v\ne 0$,
then $R_s^2$ will be smaller than the geometrical radius $R_G$. As the
transverse mass increases, this reduction of $R_{s}$ and $R_{\perp}$ relatively
to the pure geometric radius becomes more pronounced.
In the longitudinal direction, as a general consequence of the particle pair
motion with velocity $Y$, the system appears Lorentz-contracted. Hence
$q_{L}$-correlation functions at finite values of $Y$ measure longitudinal
correlation radii, which are reduced by the corresponding Lorentz-contraction
factor ${\rm ch }^{-1} Y$, as shown by (\ref{rewr}). Similar purely kinematic
factors affect the out-longitudinal radius $R_{\perp L}$.

Returning now to the higher order corrections shown in (\ref{4.8}) we observe
that, in contrast to the
nonrelativistic model, the difference between the squares of
the ``out'' and ``side'' radii depends on the rapidity $Y$ (or
$\beta_L$) of the pair and is not quite directly proportional to the
duration of particle emission $(\delta\tau)^2$ even for pairs with
$\beta_L=0$ \cite{time}. It is also worth noting that although
$R_{\perp L}^2$ vanishes when $K_\perp=0$ or $Y=0$, for high $K_\perp$
and $|Y|$ pions $R_{\perp L}^2$ is of the same order of magnitude as
$R_L^2$, so it has a significant effect on the form of the correlation
function.

This can be seen most easily in a numerical example.  For simplicity,
we consider a pion source with no transverse flow ($v=0$) which
freezes out instantaneously ($\delta\tau=0$) with the following other
source parameters: $R_G=3$ fm, $\tau_0=4$ fm/c, $\delta\eta = 1.5$,
and $T=150$ MeV.  Given these parameters and any set of momenta ${\vec
q},{\vec K}$, it is possible to determine the correlation function
both by using the approximate radii of (\ref{4.8}) and by performing
an exact numerical calculation of the correlation using (\ref{2}) and
(\ref{4.2}).  In all of our plots of the correlation function, solid
curves are used to denote numerical calculations, while dashed curves
are used to denote our analytic approximation.

\begin{figure}
\centerline{
\epsfxsize=4.5in
\epsfbox{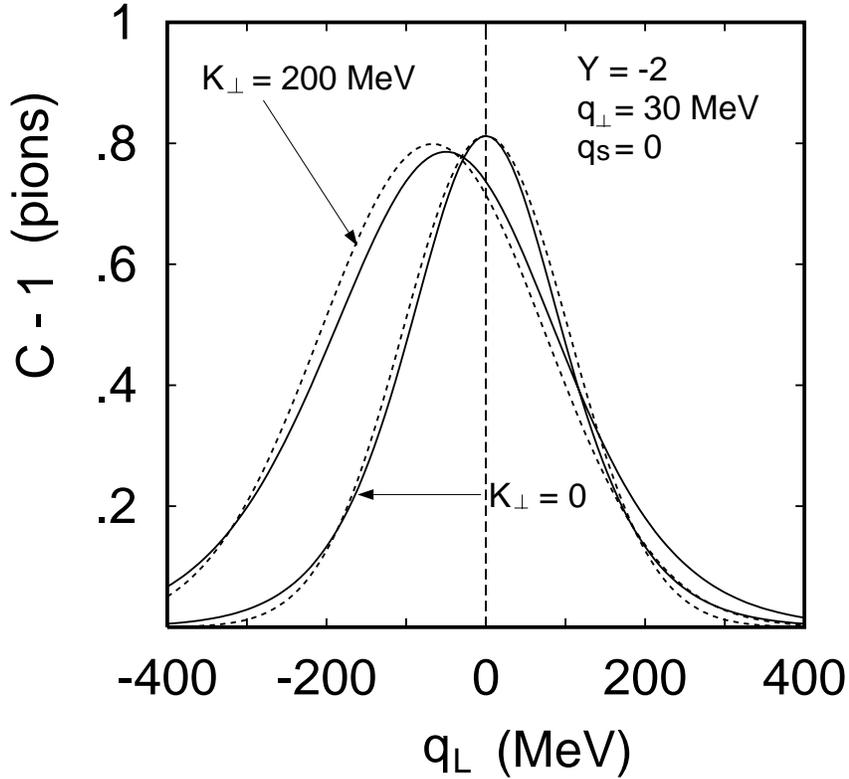}
}
\caption{The correlation function generated by the pion source of
(54) is plotted as a function of $q_L$ for two values of $K_\perp$:
$K_\perp=0$ (symmetric curves) and $K_\perp=200$ MeV (asymmetric
curves).  The solid lines are exact numerical results, while the
dashed lines are our analytic approximation (15,63).  The source
parameters are $v=\delta\tau=0$, $R_G=3$ fm, $\tau_0=4$ fm,
$\delta\eta = 1.5$, $T=150$ MeV.  The pair momenta we fixed at $Y=-2$,
$q_\perp=30$ MeV and $q_s=0$.}
\end{figure}

The symmetric curves in fig. 5 show the correlation as a function of
$q_L$ for $Y=-2$, $q_\perp=30$ MeV, $q_s=0$ and $K_\perp=0$.  In this
case as we mentioned earlier, since $K_\perp=0$ the cross term
vanishes and the correlation function peaks at $q_L=0$.  As $K_\perp$
is allowed to increase, however, $R_{\perp L}^2$ causes the peak to
shift toward negative values of $q_L$, as can be seen in the
asymmetric curves which have been calculated for $K_\perp=200$ MeV and
all of the other momenta the same.  Similarly, if $Y$ is allowed to
increase to $0$, the maximum shifts back to $q_L=0$, and for $Y>0$,
the maximum is located at a $q_L>0$.  It should also be pointed out
that in each of the above cases, the direction of the shift of the
peak is reversed if a negative $q_\perp$ is used instead of a
$q_\perp>0$.  To give a quantitative idea of how good the analytic
approximation is for the $K_\perp=200$ MeV case, we found that the
best gaussian fit to the numerical curve could be reproduced by
multiplying the correlation radii by the following factors:
$R_\perp^2\rightarrow 0.92 R_\perp^2$, $R_L^2\rightarrow 0.95 R_L^2$,
and $R_{\perp L}^2\rightarrow 0.75 R_{\perp L}^2$.

Figure 6 shows the correlation as a function of $q_\perp$ for two
different values of $q_L$.  Both sets of curves are calculated for
$Y=-2$, $q_s=0$ and $K_\perp=200$ MeV, but the upper ones have $q_L=0$
while the lower (asymmetric) ones are for $q_L=100$ MeV.  It can be
seen that increasing $q_L$ from $0$ has the effect of shifting the
peak down and to the left (to $q_\perp<0$).  This figure shows clearly
that interesting physics could be missed if correlation models are
only plotted as a function of a single momentum difference with all
other $q_i$ set equal to zero.  Again to get a quantitative idea of
the validity of the analytic approximation, we found that the best
gaussian fit to the numerical curve for $q_L=100$ MeV could be
obtained using the factors $R_\perp^2\rightarrow 0.85 R_\perp^2$,
$R_L^2\rightarrow 1.08 R_L^2$, and $R_{\perp L}^2\rightarrow 0.75
R_{\perp L}^2$.

\begin{figure}
\centerline{
\epsfxsize=4.5in
\epsfbox{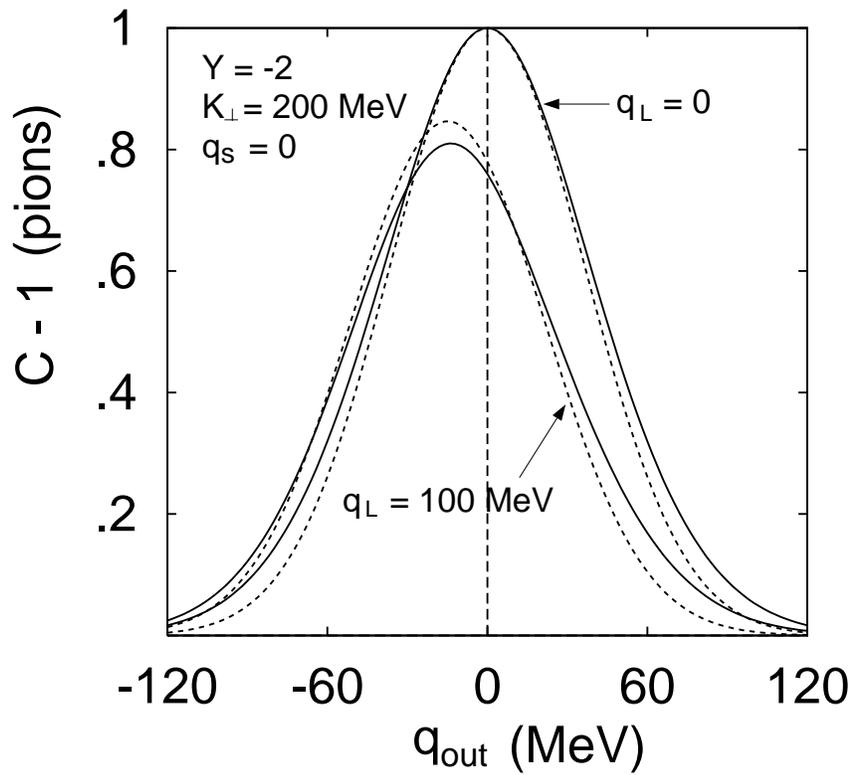}
}
\caption{The same source parameters as in fig. 5, but with pair momenta
$Y=-2$, $K_\perp=200$ MeV, $q_s=0$, are used to plot the correlation
as a function of $q_\perp$ for $q_L=0$ (symmetric curves) and
$q_L=100$ MeV (asymmetric curves)}
\end{figure}

As can be seen from figs. 5 and 6, the simple analytic expressions of
(\ref{4.8}) reproduce the exact correlation functions remarkably well
considering the crudity of the approximation.  By extensively
exploring the parameter space of the model, we have found that the
quantitative error estimates we have obtained in figs. 5 and 6 are
somewhat typical of the maximum discrepancies for reasonable
parameters.  Namely, the analytic approximations of (\ref{4.8}) for
$R_\perp^2$, $R_L^2$ and $R_{\perp L}^2$ are able to reproduce the
best gaussian fits to the numerical expressions to within
${\mathrel{\lower.9ex\hbox{$\stackrel{\displaystyle <}{\sim}$}}}
20\%$, ${\mathrel{\lower.9ex\hbox{$\stackrel{\displaystyle
<}{\sim}$}}} 10\%$,
${\mathrel{\lower.9ex\hbox{$\stackrel{\displaystyle <}{\sim}$}}}
33\%$, respectively (e.g. for $R_{\perp L}^2$, $(1-.75)/.75\sim
33\%$).  Although not shown, the analytic expressions for $R_s^2$ are
much better, their discrepancy from numerical fits is typically
${\mathrel{\lower.9ex\hbox{$\stackrel{\displaystyle <}{\sim}$}}} 5\%$.
We would also like to note that we have performed numerical
calculations using eq. (\ref{6}) and find them to agree to within 3\%
with numerical calculations using (\ref{2}), so we are well justified
in neglecting those corrections in eqs. (\ref{4.8}).

\begin{figure}
\centerline{
\epsfxsize=4.5in
\epsfbox{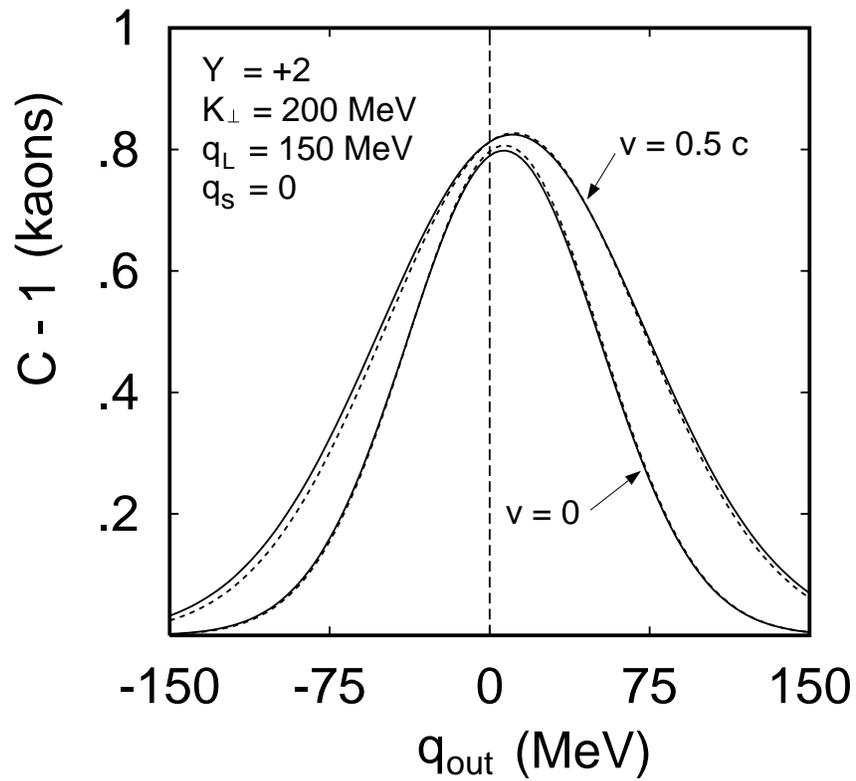}
}
\caption{The narrower curves show the kaon correlation as a function
of $q_\perp$ from a source with the same parameters as in fig. 5 but
with momenta defined by $Y=+2$, $K_\perp = 200$ MeV, $q_s=0$ and
$q_L=150$ MeV.  The wider curves have been obtained by using $v=0.5$
instead of $v=0$.}
\end{figure}

The analytic expressions of (\ref{4.8}) are even better approximations
for heavier particles like kaons, since for them $m_t/T>3$ so
$(\delta\eta)_*^2$ forms a smaller expansion parameter.  This behavior
can be seen in fig. 7 where we plot the kaon correlation as a function
of $q_\perp$ for $Y=+2$, $K_\perp = 200$ MeV, $q_s=0$ and $q_L=150$
MeV.  The narrower curves are obtained by using the same source
parameters as in figs. 5 and 6, while the wider curves feature a
transverse flow parametrized by $v=0.5c$.  The best gaussian fit to the
wider numerical curve can be obtained in this case by multiplying the
radii of the wider analytical curves by the factors
$R_\perp^2\rightarrow 0.94 R_\perp^2$, $R_L^2\rightarrow 1.02 R_L^2$,
and $R_{\perp L}^2\rightarrow 0.9 R_{\perp L}^2$.

Perhaps the best way to study the correlation function is to make a
2-dimensional surface plot of $C-1$ as function of both $q_\perp$
and $q_L$.  Figure 8 shows such a plot of the numerical calculation of
$C-1$ for $Y=-2$, $K_\perp=200$ MeV and $q_s=0$.  The effect of the
cross term can be seen in the form of a ridge running from the peak at
$q_\perp=q_L=0$ down to the front left where $q_L>0$ and $q_\perp<0$.
Since cylindrical symmetry precludes the existence of ``side-out'' or
``side-longitudinal'' cross terms, the only effect of averaging over
$q_s$ from $0$ to some maximum value such as 30 or 50 MeV
\cite{na35} would be to reduce the intercept of the
correlation function to some value less than 1.  This averaging,
however, should have very little impact on the qualitative ridge
structure of the ``out-long'' correlation function.  Consequently,
this kind of ridge should be clearly identifiable experimentally and
in fact may have already been seen in preliminary E802 correlation
data \cite{gyul2}.

\begin{figure}
\centerline{
\epsfxsize=5in
\epsfbox{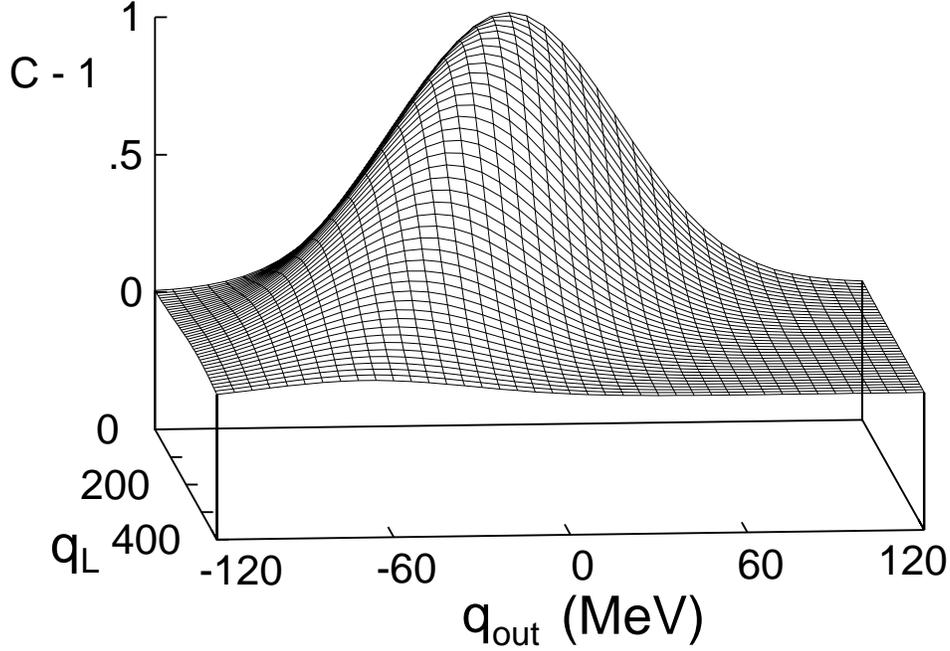}
}
\caption{The numerically calculated correlation function
generated by the pion source of fig. 5 is plotted as a function of
$q_L$ and $q_\perp$ for $Y=-2$, $K_\perp=200$ MeV, and $q_s=0$.}
\end{figure}

Before analyzing this model in rapidity coordinates, we would like to
note that the LCMS radii of this model can be obtained simply by
setting $\beta_L=0$ and $E_K=m_t$ in (\ref{4.8}). Note that the factor
of $Y$ in $R_{\perp L}^2$ should not be set equal
to zero, since it arises from the space-time rapidity distribution of the
point-like sources in (\ref{4.2}) which obviously breaks the boost invariance
of the emission function in the longitudinal direction \cite{justif}.
Transforming to the LCMS frame introduces a $Y$ dependence which
eventually translates into a nonvanishing cross term. We would like to
emphasize that, to the first order in the small expansion parameters, our
results reduce to the expressions for the LCMS correlation radii derived in
\cite{csorgo2}. However, in the light of a comparison of the results obtained
within the framework of our analytical approximation with an exact numerical
computation of the correlation function, it turns out that the second order
contributions to the correlation radii must be included. In particular, the
out-longitudinal cross terms, whose effect can be clearly seen in Fig.8, is
completely missed at leading order. Nevertheless,
for this model, the LCMS frame has the advantage that the expressions
for the correlation radii are much simpler than for those in the fixed
frame.  On the other hand, this same simplicity can be achieved without the
complication of reference frame shifting by expressing
everything in terms of boost-invariant coordinates, as we will now show.

\subsection{HBT Radii in Boost-Invariant Coordinates}
Using the model independent expressions of (\ref{2.2.10}) along with
the emission function of (\ref{4.2}), we obtain the following
correlation radii
\begin{eqnarray}
R_s^2 & = & R_*^2
\nonumber \\
R_\perp^2 & = & R_*^2 + \frac{K_\perp^2}{m_t^2}\left[
\left(1+(\delta\eta)_*^2\right)(\delta \tau)^2
+{\textstyle\frac{1}{2}}(\delta\eta)_*^4\tau_0^2\right]
\nonumber \\
\alpha^2 & = & m_t^2(\delta\eta)_*^2
\left[\tau_0^2\left(1 + \nu(\delta\eta)_*^2\right)
+(\delta\tau)^2\right]
\nonumber \\
R_{\perp {\rm y}} & = &
\frac{K_\perp Y}{(\delta\eta)^2}(\delta\eta)_*^2
\left[(\delta\eta)_*^2\tau_0^2 +(\delta\tau)^2\right]\;,
\label{4.10}
\end{eqnarray}
where in contrast to section 4.1, $Y$ is now defined
$Y={\textstyle\frac{1}{2}}({\rm y}_1+{\rm y}_2)$.  Note also that in
contrast to the corresponding radii of (\ref{4.8}) $R_\perp$ and
$\alpha$ in the above approximation are both independent of rapidity.
In addition, the cross term $R_{\perp {\rm y}}$ will be small compared
to these radii, especially for higher mass particles like kaons which
have $(\delta\eta)_*^2\ll 1$ or for future ultrarelativistic
collisions in which $(\delta\eta)\gg 1$.

The astute reader will note that aside from a difference in the
definition of $Y$, the fixed frame correlation radii of the last
subsection can be easily derived from those of (\ref{4.10}) in the
following way: First insert the radii of (\ref{4.10}) into the
expression (\ref{2.2.3}) for the correlation function, then make the
replacement ${\rm y}\rightarrow q_L/E_K-\beta_LK_\perp q_\perp/m_t^2$,
rewrite the resulting expression in the form of eq. (\ref{6g}), and
finally read off the radii of eq. (\ref{4.8}).  The reason for this
can easily be seen by noting that
\begin{eqnarray}
q{\cdot}x &\simeq & q_\perp\frac{K_\perp}{m_t}\tau\,
{\rm ch}(\eta-Y)
-{\rm y}\,m_t\,\tau\,{\rm sh}(\eta-Y) -q_\perp x -q_s y
\nonumber \\
&\simeq &q_\perp\frac{K_\perp}{m_t}\tau\,{\rm ch}(\eta-Y)
-\left(\frac{m_t}{E_K}q_L- \frac{K_\perp}{m_t}\beta_L q_\perp\right)
\,\tau\,{\rm sh}(\eta-Y)-q_\perp x -q_s y
\label{4.11}
\end{eqnarray}
where in the top line $Y={\textstyle\frac{1}{2}}({\rm y}_1+{\rm
y}_2)$, while in the bottom line
$Y={\textstyle\frac{1}{2}}\ln[(E_K+K_L)/(E_K-K_L)]$.  Note that in
particular the LCMS radii can be found simply by making the
replacement ${\rm y}\rightarrow q_L/m_t$.  Based on this equivalence,
one can see that for systems undergoing Bjorken longitudinal
expansion, {\em LCMS correlation functions are nothing more than approximations
of fixed frame
correlation functions in rapidity coordinates}.  Since the latter
formulation is manifestly boost invariant and avoids the complications arising
from the introduction of the different LCMS-reference frames, it is much more
desirable to use those coordinates.  For the
remainder of this section, we use the definition
$Y={\textstyle\frac{1}{2}}({\rm y}_1+{\rm y}_2)$.

\begin{figure}
\centerline{
\epsfxsize=5in
\epsfbox{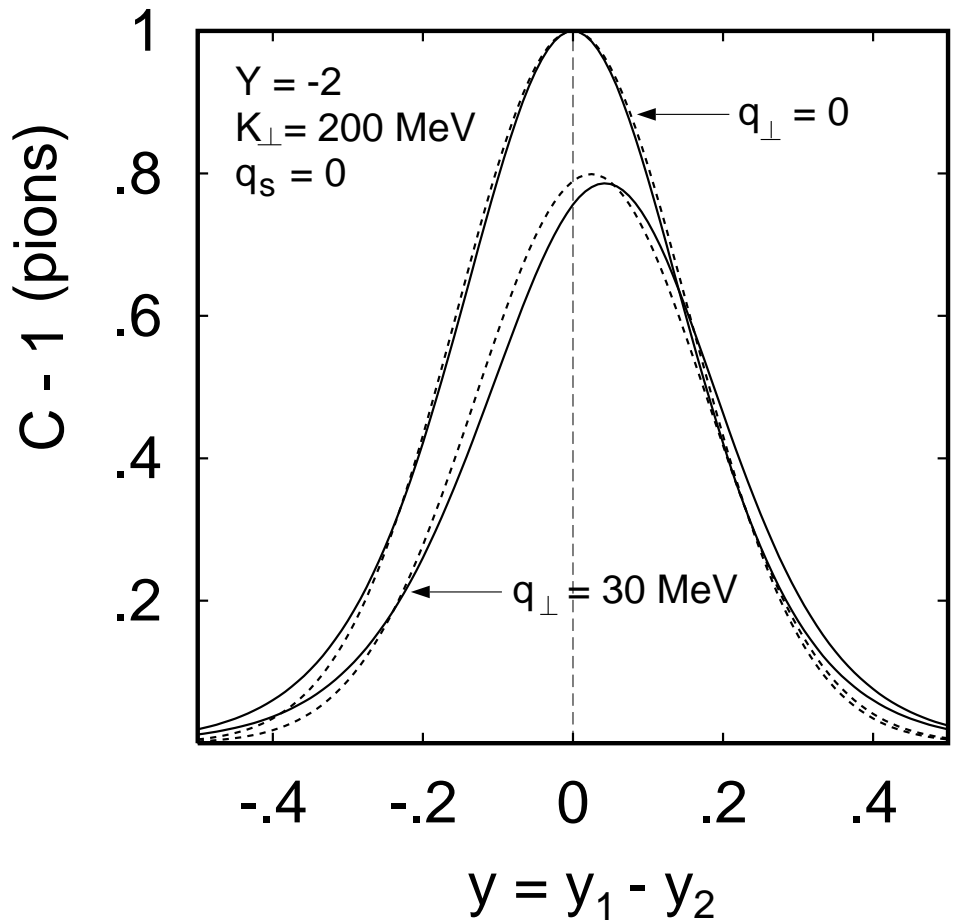}
}
\caption{The same source parameters as in fig. 5 are used to plot
the correlation as a function of ${\rm y}={\rm y}_1-{\rm y}_2$ for
$q_\perp=0$ (symmetric curves) and $q_\perp=30$ MeV (asymmetric
curves).  For both curves the pair momenta have been fixed to be
$Y=-2$, $K_\perp=200$ MeV, and $q_s=0$.}
\end{figure}

Using the same source parameters as in the last section, figure 9
shows the pion correlation as a function of ${\rm y}$ for $Y=-2$,
$K_\perp=200$ MeV and $q_s=0$.  The symmetric curves are for $q_\perp
=0$ while the asymmetric curves are for $q_\perp =30$ MeV.  The best
gaussian fit to the asymmetric numerical curve can be obtained in this
case by multiplying the radii of the wider analytical curves by the
factors $R_\perp^2\rightarrow 1.17 R_\perp^2$, $\alpha^2\rightarrow 0.96
\alpha^2$, and $R_{\perp {\rm y}}\rightarrow 1.6 R_{\perp {\rm y}}$.  Again,
these corrections are somewhat typical of the maximum discrepancies,
and the analytic approximations of (\ref{4.10}) for $R_\perp^2$,
$R_s^2$, $\alpha^2$ and $R_{\perp {\rm y}}$ are thus able to reproduce
the best gaussian fits to the numerical expressions to within
${\mathrel{\lower.9ex\hbox{$\stackrel{\displaystyle <}{\sim}$}}}
20\%$, ${\mathrel{\lower.9ex\hbox{$\stackrel{\displaystyle
<}{\sim}$}}} 5\%$, ${\mathrel{\lower.9ex\hbox{$\stackrel{\displaystyle
<}{\sim}$}}} 10\%$,
${\mathrel{\lower.9ex\hbox{$\stackrel{\displaystyle <}{\sim}$}}}
40\%$, respectively (e.g. for $R_{\perp {\rm y}}$, $0.6/1.6\sim
38\%$).  Of course, much better agreement could be obtained if a more
sophisticated analytical approximation is used in place of
eq. (\ref{4.4}).

\section{Conclusions}
By taking second derivatives of the two particle correlation function
around ${\vec q}=0$, we have derived model-independent expressions for
correlation radii both in cartesian and boost-invariant momentum
coordinates.  In both cases, an ``out-longitudinal'' cross term
arises naturally.  In the context of two ``gaussian'' models, this
term is found to have a significant effect on the form of the
correlation function.  We therefore feel that future correlation data
should be fit to one of the following two functions
\begin{equation}
C({\vec p}_1,{\vec p}_2) = 1 \pm \lambda
\exp\left(-q_s^2R_s^2 - q_\perp^2R_\perp^2 - q_L^2R_L^2 -2q_\perp q_L
R_{\perp L}^2\right)
\label{c1}
\end{equation}
or even better
\begin{equation}
C({\vec p}_1,{\vec p}_2) = 1 \pm \lambda
\exp\left(-q_s^2R_s^2 - q_\perp^2R_\perp^2 - {\rm y}^2\alpha^2
-2q_\perp\,{\rm y}\,R_{\perp {\rm y}}\right)\;,
\label{c2}
\end{equation}
where $R_{\perp L}^2$ (or $R_{\perp {\rm y}}$) can be either positive
or negative.

Currently, data is usually fit to (\ref{c1}) with $R_{\perp L}^2$ a
priori set equal to zero [14-17].  For that reason, all
$q_i$ with the same $|q_i|$ are usually binned together, but in the
process, the relative sign between $q_\perp$ and $q_L$ gets lost.
This procedure effectively averages out the cross term at the expense
of introducing large systematic errors into the measured ``out'' and
``longitudinal'' radii.  One way to avoid this averaging in practice
is to define the ordering of particles 1 and 2 by always demanding
that $q_L$ (or ${\rm y}$) be positive.  This then determines the sign
of $q_\perp$ (and $q_s$) so that positive values can be binned
separately from negative values, allowing one to generate plots like
the one we have shown in fig. 8.  Not only will measurement of
cross terms provide new information about the emitting source, it
should greatly increase the accuracy of $R_\perp$ and $R_L$ (or
$\alpha$) measurements.

The model-independent expressions for the radius parameters of the HBT
correlation function show very clearly that these parameters do not
generally measure the geometric size of the source, but rather its
lengths of homogeneity in the four space-time directions. For
expanding sources like those created in heavy ion collisions, the
gradients of the thermodynamic parameters and of the flow velocity
field contribute to the inhomogeneity of the source.  In fact, regions
of homogeneity may extend over only a small fraction of the source, in
which case the two-particle correlation function is sensitive only to
these subdomains.  Moreover, particle pairs with different average
momenta will generally see regions of homogeneity with different size,
giving rise to a characteristic ${\vec K}$-dependence of the
correlation radii.

In this paper we have studied these features quantitatively for sets
of cylindrically symmetric models with gaussian density profiles in
which the sources undergo longitudinal and transverse collective
expansion but freeze out at a constant temperature. The effect of the
flow gradients on the lengths of homogeneity and on the spatial HBT
size parameters has been seen explicitly. They lead to a reduction of
the correlation radii relative to the geometric radius parameters, and
this effect increases with the average momentum of the pair relative
to the center-of-mass of the source. The temporal length of
homogeneity of the source, given by the duration $\delta \tau$ of the
emission process, affects both the difference $R_\perp^2 - R_s^2$ (as
has been noted previously \cite{pratt,csorgo}) and the new
``out-longitudinal'' cross term. The effects of possible gradients of
the freeze-out temperature have not yet been studied in this context,
but are expected to have similar qualitative consequences.  In fact, a
difficulty in separating effects of flow gradients from those of
thermal gradients was noted before in the context of a spherically
symmetric model \cite{schnekki,mayer}.  It was found that both
mechanisms can lead to a concave curvature of the single particle
$m_t$-spectra \cite{schnekki}, as well as a similar
$K_\perp$-dependence of the ``side" and ``out" radii in the HBT
correlation function \cite{mayer}.

In \cite{csorgo,csorgo2} the difference between the geometrical
and HBT radii has been expressed in terms of a so-called ``thermal
radius". Our analysis shows that it is really not the existence of a
temperature, but of a flow velocity gradient which causes the appearance of a
length of homogeneity in the HBT radii. The temperature only plays a role as a
smearing factor, and the ratio $T/m_t$ sets the scale at which the
inhomogeneity of the flow field becomes effective. Different flow velocities in
the transverse and longitudinal directions generally lead to different
transverse and longitudinal homogeneity lengths, $R_{H}$ and $L_{H}$.
In \cite{csorgo,csorgo2} this was not obvious because the flow
gradient was fixed to be $1/\tau_0$ in all directions by the choice of
the flow velocity profile.

All of our calculations in this paper were done in a fixed reference
frame, thus avoiding the complications with the LCMS frame
discussed in Section 2. However, we found that by parametrizing the
correlation function in terms of rapidities rather than longitudinal
momenta, one finds a longitudinal correlation radius and an
out-longitudinal cross term which for sources with boost-invariant
longitudinal expansion can be well approximated by the LCMS results.
Since this parametrization avoids the LCMS problems of shifting
frames, we suggest that the concept of the LCMS be abandoned in favor
of using rapidity coordinates. We also showed that the existence of an
out-longitudinal cross term is not affected by this choice of
coordinates or frames, although its actual size is.

The analytic expressions for the HBT size parameters developed in this
paper have been tested numerically and were found to be sufficiently
accurate for being useful in obtaining good qualitative insights on
the effects which various features of the source have on the shape of
the correlation function. We also studied explicitly the usually
neglected corrections due to the off-shell nature of the average
4-momentum entering in the correlation function and found them to be
very small ($<3$\%). To the extent that our two models for the source
emission function are reasonable approximations to reality, these
relations can be used to study the effects of longitudinal and
transverse flow and of the time and duration of the freeze-out process
on the HBT data. We have checked that the models produce single
particle spectra with reasonable shapes which very likely can be used
for good fits to the data (in particular once resonance decays are
included). A more detailed analysis of the HBT data in the framework
of these models thus appears as an attractive project.

\section{Appendix}

In this appendix, we prove that in the limit $K_\perp\rightarrow 0$,
$R_\perp^2\rightarrow R_s^2$ and the cross term of either eq. (\ref{21})
or eq. (\ref{2.2.10}) vanishes (depending on the coordinate system
used).  The crucial ingredient of the proof is that the emission
function is a Lorentz scalar whose $K$ dependence only enters in the
form of scalar products with cylindrically symmetric local 4-vectors.
For simplicity in the following, we will assume that there is only one
such local 4-vector, but generalization to the emission function of
(\ref{6b}) can be done trivially.

We assume an emission function of the form
\begin{equation}
S(x,K) = \bar{S}(t,\rho,z,m^2,\psi)
\label{a1}
\end{equation}
where in a cartesian coordinate system
\begin{equation}
\psi = K{\cdot}u(x) = E_K\,u_0 - K_\perp\,u_\rho\cos\phi - K_L\,u_z
\label{a2}
\end{equation}
and $u_0$, $u_\rho$, and $u_z$ are independent of $\phi$.  Using
rapidity coordinates as in (\ref{2.2.1}) and (\ref{2.2.7}) on the
other hand, we can see that
\begin{equation}
\psi = m_t\,u_t {\rm ch}(Y-\xi) - K_\perp\,u_\rho\cos\phi
\label{a3}
\end{equation}
where $u_t$, $\xi$ and $u_\rho$ are independent of $\phi$.

In either case, as long as the $\psi$ dependence of $\bar{S}$ is
smooth, it follows that
\begin{eqnarray}
\lim_{K_\perp \to 0}
\int d^4x \cos\phi\, f(t,\rho,z)\, \bar{S} &=& 0
\nonumber \\
\lim_{K_\perp \to 0}
\int d^4x \cos\phi\, f(t,\rho,z)\, \frac{\partial\bar{S}}
{\partial\psi} &=& 0
\nonumber \\
\lim_{K_\perp \to 0}
\int d^4x \cos\phi\, f(t,\rho,z)\, \frac{\partial^2\bar{S}}
{\partial\psi^2} &=& 0
\label{a4}
\end{eqnarray}
for any $\phi$-independent function $f$.  From the first of the above
equations we can see that
\begin{equation}
\lim_{K_\perp \to 0}
\langle f(t,\rho,z)\, \cos\phi\rangle = 0
\label{a5}
\end{equation}
in particular $\langle x\rangle=\langle\rho\cos\phi\rangle=0$,
and consequently the non-derivative terms of $R_{\perp L}^2$ and
$R_{\perp {\rm y}}$ vanish in the limit $K_\perp\rightarrow 0$.
Furthermore, since
\begin{equation}
\lim_{K_\perp \to 0}
\langle \rho^2\cos^2\phi \rangle =\;
\lim_{K_\perp \to 0}
\langle \rho^2\sin^2\phi \rangle =
{\textstyle\frac{1}{2}}\langle\rho^2\rangle\;,
\label{a6}
\end{equation}
i.e. $\langle x^2\rangle = \langle y^2\rangle =
{\textstyle\frac{1}{2}}\langle\rho^2\rangle$, it can be seen that the
non-derivative terms of $R_\perp^2$ equal those of $R_s^2$ in that
limit.

As for the momentum derivative terms, from eq. (\ref{a4}) we have
\begin{equation}
\lim_{K_\perp \to 0}
\frac{d}{dK_\perp}P_1({\vec K}) = -\int d^4x\,u_\rho\cos\phi\,
\frac{\partial\bar{S}}{\partial\psi} = 0
\label{a7}
\end{equation}
in either set of coordinates.  Similarly,
\begin{equation}
\lim_{K_\perp \to 0}
\frac{d}{dK_L}\frac{d}{dK_\perp}P_1({\vec K}) =
\lim_{K_\perp \to 0}
\frac{d}{dY}\frac{d}{dK_\perp}P_1({\vec K}) = 0\;.
\label{a8}
\end{equation}
Since
\begin{equation}
\frac{d}{dK_L}\frac{d}{dK_\perp}{\rm ln}P_1({\vec K}) =
\frac{1}{P_1({\vec K})}\frac{d}{dK_L}\frac{d}{dK_\perp}P_1({\vec K}) -
\frac{1}{\left[P_1({\vec K})\right]^2}
\left(\frac{dP_1({\vec K})}{dK_L}\right)
\left(\frac{dP_1({\vec K})}{dK_\perp}\right)\;,
\label{a8a}
\end{equation}
we have proved that $R_{\perp L}^2$ of (\ref{21}) vanishes for
$K_\perp\rightarrow 0$.  The proof for $R_{\perp {\rm y}}$ follows
simply by replacing $d/dK_L$ with $d/dY$ in the above equation.

The momentum derivative term for the ``out'' radius does not vanish
in this limit, rather in the cartesian system it takes the form
\begin{equation}
\lim_{K_\perp \to 0}
\frac{d^2}{dK_\perp^2}P_1({\vec K}) =
\lim_{K_\perp \to 0}
\int d^4x\left(
\frac{u_0}{E_K}\,\frac{\partial\bar{S}}{\partial\psi}
+ u_\rho^2\,\cos^2\phi
\,\frac{\partial^2\bar{S}}{\partial\psi^2}\right)\;.
\label{a9}
\end{equation}
The derivative term for the ``side'' radius is a bit trickier since it
involves a ratio of two quantities which vanish in the
$K_\perp\rightarrow 0$ limit
\begin{equation}
\lim_{K_\perp \to 0}
\frac{1}{K_\perp}\frac{d}{dK_\perp}P_1({\vec K}) =
\lim_{K_\perp \to 0}
\int d^4x\left(\frac{u_0}{E_K} - \frac{u_\rho}{K_\perp}\cos\phi\right)
\frac{\partial\bar{S}}{\partial\psi}
\label{a10}
\end{equation}
Determination of the appropriate limit of the second term above is
found by the rule of l'Hospital by dividing the derivative (with
respect to $K_\perp$) of the numerator by the derivative of the
denominator.  When this is done, the results for the ``side'' and
``out'' directions become identical.  A similar argument can be used
to show the same thing in the rapidity coordinate system.

{\bf Acknowledgements:}

We would like to thank T. Cs\"org\"o and T. Alber for clarifying and
stimulating discussions. This work was supported in part by BMFT, DFG,
and GSI.


\begin{thebibliography}{99}
\bibitem{time} This was not seen in \cite{csorgo2} where only the first order
corrections were calculated in the LCMS frame, $\beta_{L}=0$.
\bibitem{justif}
{}From (\ref{21}) one sees that in the LCMS frame (denoted by primed variables
as in (\ref{2.7})) the cross term is given by :
\begin{eqnarray}
R_{\perp L}^2& = & \left( <x\;\tau{\rm sh}\eta'>-<x><\tau{\rm
sh}\eta'>\right)\nonumber\\
& - & \frac{K_{\perp}}{m_{t}}\left( <\tau ^{2}{\rm ch}\eta'\; {\rm sh} \eta'>-<
\tau {\rm ch} \eta'>< \tau {\rm sh}\eta'>\right)\;,\nonumber
\end{eqnarray}
up to small logarithmic corrections. This clearly vanishes as soon as the
source $S(x,K)$ becomes reflection symmetric under
$\eta'\longrightarrow-\eta'$.
Please note that for a longitudinally boost invariant velocity profile, this
symmetry is broken only by the space time rapidity distribution of the
point-like sources.
\end{thebibliography}
\end{document}